\pdfoutput=1
\documentclass[]{emulateapj}

\received{xxx}
\revised{yyy}
\accepted{zzz}

\shorttitle{Lack of Circumbinary Planets Orbiting Isolated Binaries}
\shortauthors{Fleming et al.}

\usepackage{xspace}
\usepackage{graphicx}
\usepackage{amsmath}

\def\gsim{~\rlap{$>$}{\lower 1.0ex\hbox{$\sim$}}}
\def\lsim{~\rlap{$<$}{\lower 1.0ex\hbox{$\sim$}}}

\newcommand{\vplanet}[0]{\texttt{VPLANET}\xspace}
\newcommand{\eqtide}[0]{\texttt{EQTIDE}\xspace}
\newcommand{\stellar}[0]{\texttt{STELLAR}\xspace}

\begin{document}

\title{On the Lack of Circumbinary Planets Orbiting Isolated Binary Stars}


\email{dflemin3@uw.edu}

\author{David P. Fleming}
\affil{Astronomy Department, University of Washington \\
Box 951580, Seattle, WA 98195}
\affil{NASA Astrobiology Institute - Virtual Planetary Laboratory Lead Team, USA}

\author{Rory Barnes}
\affiliation{Astronomy Department, University of Washington \\
Box 951580, Seattle, WA 98195}
\affil{NASA Astrobiology Institute - Virtual Planetary Laboratory Lead Team, USA}

\author{David E. Graham}
\affiliation{Astronomy Department, University of Washington \\
Box 951580, Seattle, WA 98195}

\author{Rodrigo Luger}
\affiliation{Astronomy Department, University of Washington \\
Box 951580, Seattle, WA 98195}
\affil{NASA Astrobiology Institute - Virtual Planetary Laboratory Lead Team, USA}

\author{Thomas R. Quinn}
\affiliation{Astronomy Department, University of Washington \\
Box 951580, Seattle, WA 98195}
\affil{NASA Astrobiology Institute - Virtual Planetary Laboratory Lead Team, USA}


\begin{abstract}


We outline a mechanism that explains the observed lack of circumbinary planets (CBPs) via coupled stellar-tidal evolution of isolated binary stars.  Tidal forces between low-mass, short-period binary stars on the pre-main sequence slow the stellar rotations, transferring rotational angular momentum to the orbit as the stars approach the tidally locked state.  This transfer increases the binary orbital period, expanding the region of dynamical instability around the binary, and destabilizing CBPs that tend to preferentially orbit just beyond the initial dynamical stability limit.  After the stars tidally lock, we find that angular momentum loss due to magnetic braking can significantly shrink the binary orbit, and hence the region of dynamical stability, over time impacting where surviving CBPs are observed relative to the boundary.  We perform simulations over a wide range of parameter space and find that the expansion of the instability region occurs for most plausible initial conditions and that in some cases, the stability semi-major axis doubles from its initial value.  We examine the dynamical and observable consequences of a CBP falling within the dynamical instability limit by running N-body simulations of circumbinary planetary systems and find that typically, at least one planet is ejected from the system.  We apply our theory to the shortest period {\it Kepler} binary that possesses a CBP, Kepler-47, and find that its existence is consistent with our model.  Under conservative assumptions, we find that coupled stellar-tidal evolution of pre-main sequence binary stars removes at least one close-in CBP in $87\%$ of multi-planet circumbinary systems.


\end{abstract}




\section{Introduction} \label{sec:intro}

To date, 11 transiting circumbinary planets (CBPs) have been discovered by {\it Kepler}.  The shortest period binary star system around which a CBP has been discovered is Kepler-47, with a binary period of 7.45 days \citep{Orosz2012}.  The lack of CBPs around shorter period binaries is probably real given the thousands of short-period ($P_{bin} \lsim 10$ days) eclipsing binaries discovered by the {\it Kepler} mission \citep{Kirk2016} and observational biases that favor their detection \citep{Munoz2015}.  From a planet formation standpoint, there should not be a severe lack of CBPs: both \citet{Alexander2012} and \citet{Vartanyan2016} show that circumbinary disks around binaries with semi-major axes $a \lsim 1$ AU provide favorable conditions for planet formation. \citet{Bromley2015} demonstrated that outside the inner region of the circumbinary disk, planet formation should occur similarly to planet formation in disks around single stars.  From these results, \citet{Bromley2015} concluded that circumbinary and single star planet occurrence rates should be similar, a claim bolstered by both \citet{Martin2014} and \citet{Armstrong2014} who find that the minimum CBP occurrence rate derived from {\it Kepler} data is of order $10\%$ and increases with CBP inclination relative to the plane of the binary.  

Although detecting CBPs via the transit method is more difficult than in the single star case \citep{Welsh2014,Winn2015}, especially since many CBPs spend less than $50 \%$ of the time in a transiting configuration \citep{Martin2017}, CBPs have in general a higher transit probability than their single star counterparts \citep{Martin2015a} making their detection feasible.  \citet{Martin2017} showed that the time-dependent chance of observing the transit of CBPs implies that continued future observations of the {\it Kepler} field could find up to 30 new CBPs, so where are the planets orbiting short-period binaries?

One explanation for the lack of transiting CBPs could simply be that most CBPs are not in a transiting configuration, perhaps due to dynamical interactions with the central binary.  However, \citet{Foucart2013} show that natal circumbinary disks, and hence the planets themselves, should be nearly coplanar with the binary due to gravitational torques from the central binary on the disk, an effect that is especially pronounced for short-period binaries.  Furthermore, in an analysis of the observed population of {\it Kepler} CBPs, \citet{Li2016} find that the observed coplanarity of CBPs and their host binaries is not due to a selection bias.  From both theoretical arguments and analysis of {\it Kepler} data, it seems that an additional physical mechanism is required to explain the lack of discovered transiting CBPs in the {\it Kepler} field around short-period binaries.

Several recent studies have invoked the presence of a stellar tertiary companion to explain not only how short-period binaries could form but also to explain the lack of CBPs around short-period binaries. \citet{Fabrycky2007} showed that secular interactions with a tertiary companion can drive Kozai-like oscillations that cause large eccentricity oscillations in the inner binary.  The increased binary eccentricity leads to efficient tidal dissipation in the inner binary, shrinking the orbital period to of order a day from much longer periods. The comprehensive population synthesis study by \citet{Moe2018} support this finding and show that the combination of tidal dissipation and Kozai-like oscillations due to a tertiary companion can account for ${\sim}40\%$ of binaries with periods $\lsim 10$ days.  \citet{Munoz2015}, \citet{Martin2015b}, and \citet{Hamers2016} all show that these binary-tertiary interactions, in addition to shrinking the inner binary orbit, can lead to rich dynamical interactions that can drive many CBPs towards eccentric and inclined orbits, making detection more difficult and potentially leading to orbital instability. This mechanism provides a particularly compelling explanation for the lack of CBPs around short-period binaries given that in a survey of solar-type binaries, \citet{Tokovinin2006} find that $96\%$ of binaries with periods $\lsim 3$ days have a tertiary companion.  However, no study to date has examined the lack of CBPs around isolated binaries, i.e., binaries without a tertiary companion.  Not all close binaries have a companion, as \citet{Tokovinin2006} find that the tertiary companion fraction decreases to $34\%$ for binaries with periods $\gsim 12$ days after correcting for observational biases, indicating that binaries with orbital periods $\gsim 3$ days are less likely to have a tertiary companion, and therefore the Kozai-like oscillations model cannot solely account for their lack of observed CBPs.

Short-period isolated binaries can form through a combination of fragmentation and dynamical processing. \citet{Bonnell1994} found that very low mass ($\lsim 0.01$M$_{\odot}$) binaries with separations $\lsim 1$AU can form either in a protoplanetary disk orbiting an unstable protostellar core or from the unstable core itself, and must accrete mass to become a stellar binary.  When close binaries do form, simulations by \citet{Bate2000} find that they are likely to host circumbinary disks, necessary for CBP formation.  Circumbinary disks play a major role in hardening the central binary and increasing its mass; \citet{Bate2000} shows that accretion from a circumbinary disk is likely to shrink the binary separation, sometimes by up to 2 orders of magnitude.  Simulations by \citet{Arty1996} find that binaries can efficiently accrete mass from a circumbinary disk, indicating that shrinking the binary semi-major axis via accretion can readily occur.  With the inclusion of a realistic treatment of magnetic fields in MHD simulations of protobinary stars, \citet{Zhao2013} find that binary orbital decay via accretion is significantly enhanced relative to simulations without magnetic fields.  Additionally, gravitational torques between a circumbinary disk and the central binary shrink the binary semi-major axis \citep[e.g.][]{Arty1991,Bate2002,Armitage2005,Fleming2017}, which, when coupled with accretion, can produce short-period, isolated binaries.

\begin{figure}[]
	\includegraphics[width=\columnwidth]{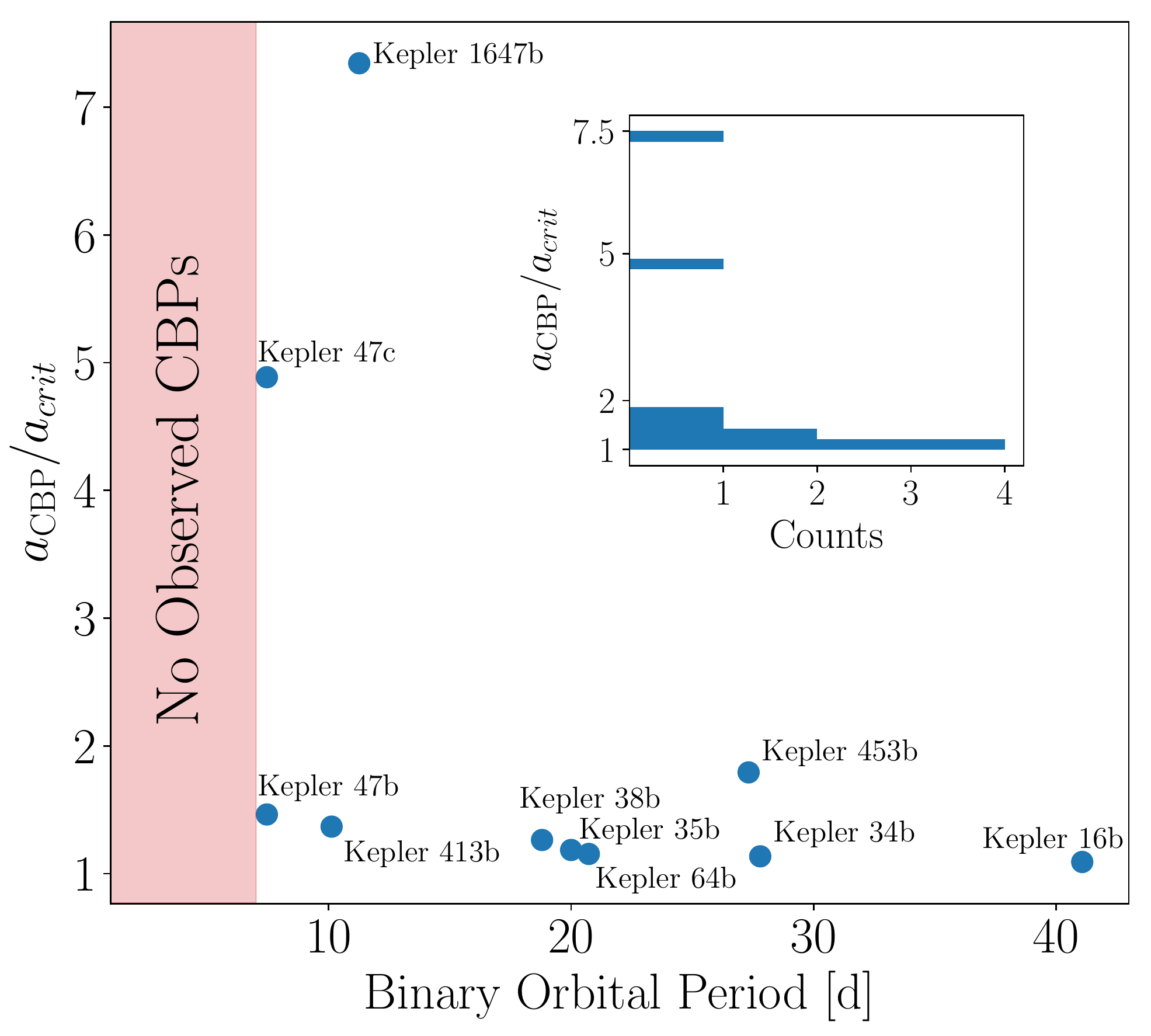}
    \caption{Semi-major axes of confirmed {\it Kepler} CBPs relative to the host binary's critical semi-major axis ($a_{\text{CBP}}/a_{crit}$) as a function of binary orbital period. Each point is annotated with the planet's name.  The red-shaded region highlights the observational finding that, to date, no CBPs have been discovered orbiting binaries with binary orbital periods less than 7.45 days. \citep{Welsh2014,Winn2015}.  We neglect Kepler-47d as its orbital parameters are not well constrained. {\it Inset:} Histogram of observed {\it Kepler} CBPs' $a_{\text{CBP}}/a_{crit}$.  The histogram demonstrates that most CBPs' semi-major axes cluster near the dynamical stability limit, $a_{crit}$.}
    \label{fig:observed_acrit}
\end{figure}

In this paper, we focus on isolated binaries with binary orbital periods in the regime $3 \leq P_{bin} \leq 7.45 $ days, as these binaries are less likely to have a tertiary companion than binaries with $P_{orb} \leq 3$ days \citep{Tokovinin2006}.  The upper limit of this range corresponds to the orbital period of Kepler-47, the shortest period CBP-hosting binary system.  We also consider the full population of \textit{Kepler} CBPs.  As previously mentioned, in the {\it Kepler} sample there are no known CBPs orbiting the ${\sim 2000}$ eclipsing binaries with $P_{orb} \lsim 7.45$ days \citep{Kirk2016}, highlighted in the red-shaded region in Fig.~\ref{fig:observed_acrit}.  Note that CBPs have been discovered by other means, such as microlensing \citep[e.g.][]{Bennett2016}. 


 
One intriguing characteristic of the observed population of {\it Kepler} CBPs is their tendency to orbit just exterior to the dynamical stability limit \citep{Welsh2014,Winn2015}.  The dynamical stability limit, referred to as the ``critical semi-major axis" ($a_{crit}$), is the minimum semi-major axis for a CBP to remain dynamically stable \citep{Dvorak1989,Holman1999}.  \citet{Holman1999} derived and empirical formula for $a_{crit}$ from an ensemble of N-body simulations given by
\begin{equation}
\begin{split} \label{eqn:crit_semi}
a_{crit} = & (1.60 + 5.1e -2.22 e^2 + 4.21 \mu \\
& -4.27e \mu -5.09 \mu^2 + 4.61 e^2 \mu^2) a \\
\end{split}
\end{equation}
where $a$ is the binary semi-major axis, $e$ the binary eccentricity and $\mu = m_2 / (m_1 + m_2)$ is the binary mass ratio.  We note that the $a_{crit}$ derived by \citet{Holman1999} is not a hard limit as Eq.~(\ref{eqn:crit_semi}) has an error of about $3\%$ - $6\%$. We plot the semi-major axis of observed {\it Kepler} CBPs normalized by $a_{crit}$ ($a_{CBP}/a_{crit}$) as a function of host binary orbital period in Fig.~\ref{fig:observed_acrit} to demonstrate CBPs' tendency to orbit just exterior to $a_{crit}$.  Using Nbody simulations, \citet{Quarles2018} found that some {\it Kepler} circumbinary systems could host an additional planet interior to the observed one.  Clearly, however, the observed CBPs cluster near the stability limit as the distribution of the ratio of $a_{CBP}$ to their host binary's $a_{crit}$ has a minimum of $a_{cbp}/a_{crit}{\approx} 1.1$ and a median of $a_{cbp}/a_{crit}{\approx} 1.26$. Analyses by \citet{Martin2014} and \citet{Li2016} show that this clustering does not solely stem from an observational bias, suggestion a physical origin.

CBP migration in a protoplanetary disk provides a compelling physical explanation for the observed pile-up of planets near the dynamical stability limit around binary stars.  Numerous studies of planet formation in circumbinary disks show that CBPs likely did not form {\it in situ} \citep[e.g.][]{Paardekooper2012,Meschiari2012a,Meschiari2012b,Pelupessy2013} and instead migrated to their present location.  Simulations of CBPs embedded in a natal protoplanetary disk by \citet{PierensNelson2007} show that the planets migrate inward, halting in the region of stability just outside of the central disk cavity that is formed from binary gravitational truncation \citep{Arty1994}.  \citet{Dunhill2013}, \citet{Pierens2013}, and \citet{Kley2014} all find that CBP migration can explain the observed properties of CBPs discovered by {\it Kepler}, depending on the precise mass and structure of the disk, suggesting that the pile-up of CBPs near the dynamical stability limit is indeed a real and expected effect.  If these planets did in fact form farther out in the disk and migrate inward, they must have formed in the first few Myrs of the system's existence before the disk dispersed, given typical lifetimes of protoplanetary disks \citep[e.g.][]{Haisch2001}.

Whether due to migration or uncharacterized circumbinary disk physics, the pile-up of CBPs near $a_{crit}$ seems to have a physical origin.  The other important characteristic of the observed {\it Kepler} CBP population, the binary orbital period below which no CBPs are observed, $7.45$ days (see Fig.~\ref{fig:observed_acrit}), likely has a physical origin, as well.  Curiously, this cutoff is within the range of binary orbital periods \citet{Zahn1989} found that separates eccentric from circular binary systems, $7.2-8.5$ days.  The theoretical cutoff identified by \citet{Zahn1989} in their study of coupled stellar-tidal physics effectively characterizes the approximate binary orbital period at which the influence of tides becomes important to the system's evolution.  In this case, it seems that in addition to stellar evolution, tidal processes in binary star systems may impact the observed circumbinary planet distribution.

In this work, we propose that the lack of CBPs around short-period binary stars is a natural outcome of coupled stellar-tidal binary evolution that we describe as follows:  In young binary systems, tidal forces synchronize the stellar rotations to match the orbital period, transferring rotational angular momentum to the orbit, increasing the orbital semi-major axis, and finally, expanding the region of dynamical instability around the binary and engulfing CBPs.  CBPs located just exterior to the initial dynamical stability limit enter the expanding instability region, become destabilized, and can be ejected from the system.  We refer to this proposed mechanism, the Stellar-Tidal Evolution Ejection of Planets, as the STEEP process for notational convenience.  

In $\S$~\ref{sec:methods}, we detail our computational methods and outline the mathematics of our theory.  We outline our experimental scheme in $\S$~\ref{sec:simulations} and discuss the results of our simulations in $\S$~\ref{sec:results} and $\S$~\ref{sec:nbody_results}.  We apply our theory to the Kepler-47 system in $\S$~\ref{sec:kepler47} and explore the implications our results and future prospects in $\S$~\ref{sec:discussion}.

\section{Methods} \label{sec:methods}

In this section, we outline how we simulate coupled stellar-tidal evolution using the code \vplanet \citep[][Barnes \textit{et al.}, \textit{in prep}]{Barnes2016} and perform N-body simulations using the code \texttt{REBOUND} \citep{Rein2012} to probe the stability of circumbinary planetary systems in which the inner-most planet falls within $a_{crit}$ as a result of coupled stellar-tidal evolution.


\subsection{\vplanet} \label{sec:vplanet}

We simulate coupled stellar-tidal binary star evolution using the code \vplanet, a modular code that allows the user to specify which physical processes impact a given variable.  Each physical process, here referred to as a module, is given by a set of nonlinear ordinary differential equations or explicit functions of time (see $\S$~\ref{sec:stellar_evolution} for the \stellar module, $\S$~\ref{sec:tidal_evolution} for the \eqtide module, and $\S$~\ref{sec:coupled_evolution} for additional coupling of \stellar and \eqtide for an in-depth description and their respective equations).  \vplanet provides a framework in which equations from different modules are coupled such that different physical processes impact the evolution of a given variable by summing the time derivatives from each relevant module.  For $n$ modules impacting the evolution of the variable $x$, at each timestep \vplanet computes the time derivative of $x$ as
\begin{equation} \label{eqn:vplanet_dxdt}
\left( \frac{dx}{dt} \right)_{tot} = \left( \frac{dx}{dt} \right)_{1} + \left( \frac{dx}{dt} \right)_{2} + ... + \left( \frac{dx}{dt} \right)_{n}.
\end{equation}
This numerical setup allows \vplanet to simultaneously integrate an arbitrary number of coupled nonlinear ODEs to rapidly simulate a system in which numerous physical processes impact the system, such as tidally interacting pre-main sequence binaries.

\vplanet numerically integrates the equations using a fourth-order Runge-Kutta scheme with adaptive time-stepping.  Our time-stepping algorithm chooses the timestep to resolve the evolution of the fastest changing variable for each simulation step in order to ensure that we completely resolve the evolution of the system.  The timescale over which a given variable $x$ changes is estimated by computing $|x|/|dx/dt|$ where $dx/dt$ is the instantaneous derivative of the variable $x$ computed by \vplanet via Eq.~(\ref{eqn:vplanet_dxdt}).  Each simulation step, we calculate the timestep by computing, for each variable, its evolutionary timescale under each module and multiply the minimum value by a scale factor, $\eta$.   We find that our simulations converge and approximately conserve both energy and angular momentum to ${\sim}10^{-4}$ when $\eta \lsim 10^{-3}$ (see $\S$~\ref{sec:conservation} and $\S$~\ref{sec:fiducial_simulation}).



\subsection{Stellar Evolution} \label{sec:stellar_evolution}

In our simulations, we track how a star's rotation rate and radius change over time due to stellar evolution using a module called \stellar.  The two main stellar evolution processes that impact a star's radius and rotation rate are stellar contraction/expansion and magnetic braking.  In general, a star's radius will contract during the pre-main sequence phase and expand slowly during the main sequence.  Conservation of angular momentum dictates that as a star contracts, its rotation rate increases (and vice-versa).  We model a star's radius as a function of time using a cubic spline interpolation of the radius tracks for a star of a given mass from the stellar evolution models of \citet{Baraffe2015} for solar metallicity stars.

We derive the time derivative of a star's rotation rate due to both magnetic braking and radius evolution under conservation of angular momentum.  For simplicity, we model a star as a solid body with a given density profile parameterized by the radius of gyration, $r_g$, where the moment of inertia is given by $I = m r_g^2 R^2$ for mass $m$ and radius $R$.  We assume solid body rotation for stars as the surface rotation evolution of low-mass (${\lsim}1$M$_{\odot}$) stars can be reasonably approximated by assuming stellar solid-body rotation \citep{Bouvier1997} and since adopting stellar solid-body rotation is common amongst studies examining stellar-tidal interactions \citep[e.g.][]{DobbsDixon2004,Heller2011,Barnes2013,Repetto2014,Bolmont2016,Bolmont2017}.  We neglect effects such as differential rotation and changes in $r_g$ but perform sensitivity tests on $r_g$ in $\S$~\ref{sec:var_rg}.

\begin{figure*}[t]
	\includegraphics[width=\textwidth]{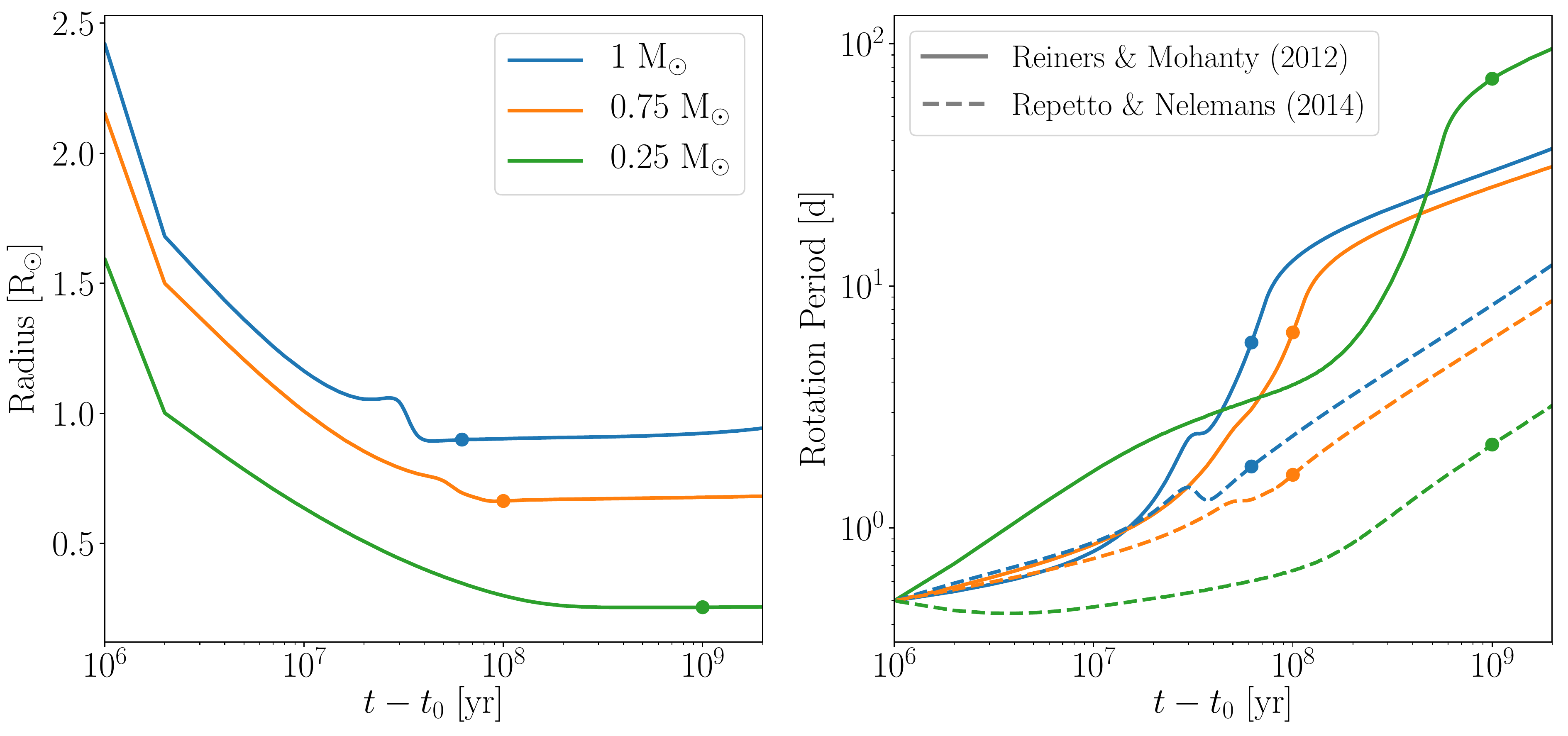}
    \caption{Stellar radius and rotation period evolution as computed by our stellar evolution model, \stellar, for a 1 M$_{\odot}$ (blue), 0.75 M$_{\odot}$ (orange), and 0.25 M$_{\odot}$ (green) star both with solar metallicity. The dots in both panels indicate the approximate time each star reaches the Zero Age Main Sequence.  {\it Left:} Stellar radius as a function of time according to a cubic spline interpolation of the \citet{Baraffe2015} stellar models.  {\it Right:} Stellar rotation period as a function of time computed via stellar radius evolution and both magnetic braking models under conservation of angular momentum (see Eq.~(\ref{eqn:stellar_rot_rate_dt})).  The solid lines correspond to simulations using the \citet{Reiners2012} magnetic braking law while the dashed lines use the \citet{Repetto2014} magnetic braking law.}
    \label{fig:stellar_example}
\end{figure*}

The rotational angular momentum for a star is simply $J = I \omega$, where $\omega$ is the rotational frequency.  By conservation of angular momentum, the star's rotation rate changes due to stellar radius evolution according to
\begin{equation} \label{eqn:rot_ang_mom_dt}
\dot{\omega}_{contraction} = \frac{-2 \dot{R} \omega}{R}.
\end{equation}
A star loses angular momentum due to magnetic braking, decreasing $\omega$.  Magnetic braking is caused by the corotation of the stellar wind with the star's magnetic field lines \citep[see][]{Parker1958,Mestel1968}.  The poloidal magnetic field of the star carries the corotating mass far away from the star, effectively removing angular momentum from the star.  Even though mass loss rates for sun-like stars are small \citep[e.g., $\dot{M}{\sim}10^{-14}$ M$_{\odot}$/yr;][]{Tarduno2014}, a star's rotation rate can slow appreciably with time due to this effect (see Fig.~\ref{fig:stellar_example}).  

Numerous models for stellar magnetic braking have been examined in the literature and here we consider two models.  The first is from \citet{Reiners2012}, who derived their model in the context of relating stellar rotation to stellar magnetic field strength.  The model is calibrated to reproduce observations of the Sun's current rotation period and the rotation-mass distribution of field stars that are a few Gyr-old.  \citet{Reiners2012} give the change in stellar angular momentum due to magnetic braking as
\begin{equation} \label{eqn:ang_mom_loss_reiners}
\begin{split}
\frac{dJ_{\star}}{dt} & = -C \left[ \omega \left(\frac{R^{16}}{m^2} \right)^{1/3} \right] \text{for $\omega \geq \omega_{crit}$} \\
\frac{dJ_{\star}}{dt} & = -C \left[ \left( \frac{\omega}{\omega_{crit}} \right)^4 \omega \left(\frac{R^{16}}{m^2} \right)^{1/3} \right] \text{for $\omega < \omega_{crit}$},
\end{split}
\end{equation}
where the authors find a best fit of $C = 2.66 \times 10^3 \text{ (gm$^5$ cm$^{-10}$ s$^3$)$^{1/3}$}$, $\omega_{crit} = 8.56 \times 10^{-6}\text{ s$^{-1}$}$ for $m > 0.35$ M$_{\odot}$, and $\omega_{crit} = 1.82 \times 10^{-6} \text{ s$^{-1}$}$ for $m \leq 0.35 M_{\odot}$.

The second magnetic braking model we consider is presented in \citet{Repetto2014} and is derived from the empirical relation for stellar spin-down of Sun-like stars empirically derived by \citet{Skumanich1972}.  The change in angular momentum due to this spin-down law is given by
\begin{equation} \label{eqn:ang_mom_loss_skumanich}
\frac{dJ_{\star}}{dt} = - \gamma m r_g^2 R^4 \omega^3
\end{equation}
where $\gamma = 5 \times 10^{-25} \text{ s m}^{-2}$ \citep{Repetto2014}.  Assuming one of the magnetic braking laws for $\dot{J}_{\star}$, the change in stellar rotation rate due to magnetic braking is
\begin{equation} \label{eqn:rot_rate_dt}
\dot{\omega}_{MB} = \frac{\dot{J}_{\star}}{I}
\end{equation}
for a fixed stellar radius under conservation of angular momentum assuming negligible mass loss.  

Both magnetic braking laws presented above are derived for spin-down rates of single stars, while in this work, we apply them to the evolution of short-period stellar binaries.  These magnetic braking laws, however, have successfully been used to model the evolution of short-period binary systems ranging from compact object-stellar binaries \citep[e.g.][]{Verbunt1981,Repetto2014}, the formation of main sequence stellar contact binaries \citep[e.g.][]{Stepien1995,Andronov2006}, and cataclysmic variable evolution \citep[e.g.][]{Ivanova2003}.  Therefore, our usage of magnetic braking laws derived for single stars is valid in this context.

We combine Eq.~(\ref{eqn:rot_ang_mom_dt}) and Eq.~(\ref{eqn:rot_rate_dt}) to get the net change in stellar rotation rate due to magnetic braking and stellar radius evolution under conservation of angular momentum:
\begin{equation} \label{eqn:stellar_rot_rate_dt}
\dot{\omega} = \dot{\omega}_{contraction} + \dot{\omega}_{MB} = \frac{\dot{J}_{\star}}{I} - \frac{2 \dot{R} \omega}{R}.
\end{equation}
and $\dot{J}$ is given by either Eq.~(\ref{eqn:ang_mom_loss_reiners}) or Eq.~(\ref{eqn:ang_mom_loss_skumanich}).

In Fig.~\ref{fig:stellar_example}, we plot the stellar radius and rotation period evolution for solar metallicity low-mass stars assuming $r_g = 0.27$ using both magnetic braking laws to demonstrate the qualitative behavior of our stellar evolution model, \stellar.  In general, the stellar radii contract along the pre-main sequence and slowly expand on the main sequence, while the stellar rotations slow over time due to magnetic braking.  


\subsection{Tidal Evolution} \label{sec:tidal_evolution}

For our tidal physics, we use a variant of the equilibrium tidal theory first introduced by \citet{Darwin1880}, the ``Constant Phase Lag" (CPL) equilibrium tide theory as derived in \citet{FerrazMello2008} in our module, \eqtide.  Equilibrium tidal theories predict that gravitational torques between the bodies and their respective tidal bulges drive a secular evolution in the eccentricity ($e$), semi-major axis ($a$) and the bodies' spins ($\omega$) and obliquities ($\psi$).  The CPL model assumes the tidally-interacting bodies raise tidal bulges on each other that maintain a fixed phase with respect to the line connecting the bodies' centers of mass.  A tidal bulge is composed of a linear sum of discrete tidal lags each with their own respective frequency.  Each tidal lag's frequency is independent of any orbital or rotational forcing frequency and there is no coupling between tidal lags.  In this formalism, the CPL model is akin to a driven, damped harmonic oscillator \citep{Greenberg2009}.  This theory, accurate up to second order in $e$, has been used extensively in many previous studies \citep[e.g.][]{Leconte2010,Heller2011,Barnes2013} and has successfully reproduced the qualitative tidal evolution of Solar System bodies \citep[e.g.][]{Goldreich1966}.  Given the physical complexity of tidally interacting astrophysical bodies, linear equilibrium tidal models, such as the CPL model, are likely not valid at large $e$ or inclinations as the linearity assumption breaks down \citep{FerrazMello2008,Greenberg2009}.  To maintain qualitative accuracy in our tidal evolution, we restrict the binary eccentricity to $e \lsim 0.2$.  We note that other equilibrium tidal theories exist, such as the ``Constant Time Lag" (CTL) model \citep[e.g.][]{Hut1981}, but we do not consider them here since in our adopted eccentricity regime, \citet{Leconte2010} showed that both the CPL and CTL model yield similar results.

Below we present a form of the CPL model tidal evolution given by \citet{Heller2011} with a modification for synchronous rotators from \citet{FerrazMello2008}.  Note that in this work we set all obliquities to 0 and do not consider their evolution, however we include it in our model for completeness.  The CPL equations for $e$ and $a$ evolution are:
\begin{equation} \label{eqn:cpl_dedt}
\frac{de}{dt} = -\frac{ae}{8 G m_1 m_2} \sum_{i=1}^2 Z_i^{'} \left( 2 \varepsilon_{0,i} - \frac{49}{2} \varepsilon_{1,i} + \frac{1}{2} \varepsilon_{2,i} + 3 \varepsilon_{5,i} \right)
\end{equation}
\begin{equation} \label{eqn:cpl_dadt_net}
\frac{da}{dt} = \sum_{i=1}^2 \frac{da_i}{dt}
\end{equation}
where if the $i^{th}$ body is tidally locked in a synchronous orbit,
\begin{equation} \label{eqn:cpl_dadt_locked}
\frac{da_{i,sync}}{dt} = -\frac{a^2}{G m_1 m_2} Z_i^{'} \left( 7 e^2 + \sin^2 (\psi_i) \right) \varepsilon_{2,i},
\end{equation}
otherwise
\begin{equation}
\begin{split}
\frac{da_i}{dt} & = \frac{a^2}{4 G m_1 m_2} Z_i^{'} \left( 4 \varepsilon_{0,i} + e^2 \left[ -20 \varepsilon_{0,i} + \frac{147}{2} \varepsilon_{1,i} \right. \right. \\
&  + \left. \left. \frac{1}{2} \varepsilon_{2,i} - 3 \varepsilon_{5,i} \right] - 4 \sin^2 (\psi_i) \left[ \varepsilon_{0,i} - \varepsilon_{8,i} \right] \right).
\end{split}
\end{equation}
The CPL equations for $\psi$ and $\omega$ evolution are
\begin{equation} \label{eqn:cpl_dpsidt}
\frac{d\psi_i}{dt} = \frac{Z_i^{'} \sin(\psi_i)}{4 m_i r_{g,i}^2 R_i^2 n \omega_i} \left( [1-\xi_i] \varepsilon_{0,i} + [1+\xi_i](\varepsilon_{8,i} - \varepsilon_{9,i}) \right)
\end{equation}
\begin{equation} \label{eqn:cpl_dwdt}
\begin{split}
\frac{d\omega_i}{dt}& = -\frac{Z_i^{'}}{8m_i r_{g,i}^2 R_i^2 n} \left(4 \varepsilon_{0,i} + e^2\left[-20\varepsilon_{0,i} + 49\varepsilon_{1,i} + \varepsilon_{2,i} \right] \right. \\
& \left. + 2 \sin^2(\psi_i) \left[ -2 \varepsilon_{0,i} + \varepsilon_{8,i} + \varepsilon_{9,i} \right] \right)
\end{split}
\end{equation}
for the $i^{th}$ body where $G$ is Newton's gravitational constant, $n$ is the binary's mean motion , and $\varepsilon$ denote the signs of the tidal phase lags.

The intermediate variables $Z_i^{'}$ and $\xi_i$ are given by
\begin{equation} \label{eqn:cpl_z}
Z_i^{'} = 3 G^2 k_{2,i} m_j^2 (m_1 + m_2) \frac{R_i^5}{a^9} \frac{1}{n Q_i}
\end{equation}
\begin{equation} \label{eqn:cpl_xi}
\xi_i = \frac{r_{g,i}^2 R_i^2 \omega_i a n}{G m_j}
\end{equation}
where the $j^{th}$ body is the $i^{th}$ body's companion in the binary, $k_{2,i}$ is the $i^{th}$ body's Love number of degree 2, and $Q$ is the tidal quality factor (also referred to as the ``tidal Q").  For all stars in all simulations, we assume $k_2 = 0.5$.  This choice of $k_2$ does not impact our results as it is degenerate with the choice of tidal Q via the $k_2/Q$ scaling in Eq.~(\ref{eqn:cpl_z}).  We choose to vary stellar tidal Qs to probe how different tidal dissipation rates impact our results (see $\S$~\ref{sec:var_Q}).

The signs of the tidal phase lags for the $i^{th}$ body are given by
\begin{equation} \label{eqn:cpl_eps}
\begin{split}
\varepsilon_{0,i} & = \Sigma(2 \omega_i - 2n) \\
\varepsilon_{1,i} & = \Sigma(2 \omega_i - 3n) \\
\varepsilon_{2,i} & = \Sigma(2 \omega_i - n) \\
\varepsilon_{5,i} & = \Sigma(n) \\
\varepsilon_{8,i} & = \Sigma(\omega_i - 2n) \\
\varepsilon_{9,i} & = \Sigma(\omega_i)
\end{split}
\end{equation}
where $\Sigma(x)$ gives returns $1$ for positive $x$, $-1$ for negative $x$, or $0$ otherwise.

As a system approaches a tidally locked state, the numerical integration of our tidal equations can become unstable due to the discrete nature of the CPL model and of the integration scheme itself.  For example, if a simulation approaches a tidally locked and synchronous state, $\omega \approx n$, then the derivatives of the tidal equations become discontinuous.  Numerical integration is inherently a discrete scheme, so solutions for systems near such a state will oscillate around a 1:1 spin-orbit resonance, causing $\varepsilon_{0,i}$ to rapidly switch signs and hence change how the systems evolve, leading to unstable, unphysical behavior.  To rectify this issue, once a body's spin period is within $1\%$ of the orbital period, we force the system into a tidally locked, synchronous state by setting $\omega = n$, following \citet{Barnes2013}.  In $\S$~\ref{sec:coupled_evolution}, we derive equations that account for the coupled stellar-tidal evolution for tidally locked star(s) to conserve both energy and angular momentum and ensure that our model results in a physically realistic evolution.

Stars on eccentric orbits can enter a pseudo-synchronous rotation state or become trapped in a higher order spin orbit resonance when the system tidally locks, with a familiar example being Mercury's 3:2 spin-orbit resonance \citep{GoldreichPeale1966}.  Note that here the use of ``spin-orbit resonance" does not mean this system is trapped in a dynamical resonance in the traditional sense, but instead enters into a spin-orbit commensurability in which the spin and rotational frequencies are integer multiples of each other; we use ``spin-orbit resonance" for notational convenience.  If a body tidally locks into a pseudo-synchronous rotation state, the rotational period is a continuous function of both the orbital eccentricity and period \citep[see][]{Goldreich1966b,Wisdom2008}.  The CPL model, however, only permits 2 rotation states for tidally locked bodies, a 1:1 and 3:2 spin-orbit resonance \citep{Barnes2017}.  In the CPL model an orbit is trapped in a 3:2 spin-orbit resonance when the binary tidally locks with $e \geq \sqrt{1/19} \approx 0.229$ \citep{FerrazMello2008} and locks into synchronous rotation otherwise.  In our adopted eccentricity regime, $e \lsim 0.2$, stars tidally lock into a synchronous state with $\omega = n$ but in $\S$~\ref{sec:32} we probe how capture into a 3:2 spin-orbit-resonance for more eccentric binary star systems impacts our results. 

\begin{figure*}[t]
	\includegraphics[width=\textwidth]{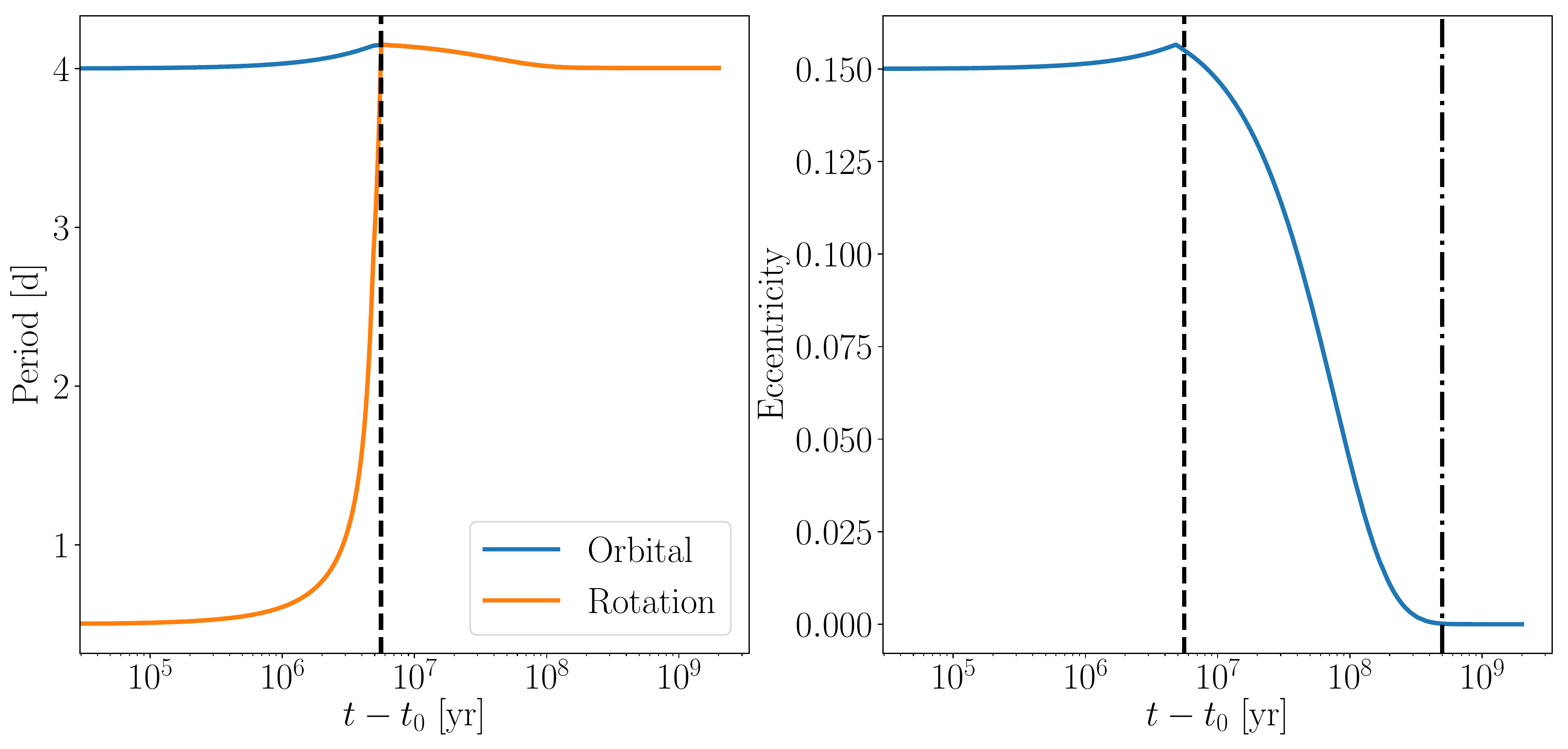}
    \caption{Tidal evolution of a binary star system with no stellar evolution under the CPL model \citep{FerrazMello2008,Heller2011} as computed by our tidal evolution model, \eqtide.  See the text for the system properties. {\it Left:} Binary orbital and stellar rotational period versus time.  Tides transport angular momentum from the stellar rotations into the orbit until the binary becomes tidally locked and synchronous (denoted by the black dashed line).  {\it Right:} Binary orbital eccentricity versus time.  After the orbit synchronizes, the eccentricity decreases until the binary circularizes (denoted by the dash-dotted line).}
    \label{fig:eqtide_example}
\end{figure*}

In Fig.~\ref{fig:eqtide_example}, we plot the tidal evolution of a $1$ M$_{\odot}$$-1$ M$_{\odot}$ binary star system using the fiducial parameters given in Table~\ref{tab:params} to demonstrate the qualitative behavior of our tidal evolution model, \eqtide.  For details of the numerical intergration of the simulation, see $\S$~\ref{sec:vplanet}.  In general, tides transfer angular momentum from the stellar rotations into the orbit until the binary reaches a tidally locked, synchronous orbit after ${\sim}10^6$ yr.  The binary orbit circularizes after ${\sim} 5 \times 10^8$ yr.

\subsection{Coupled Stellar-Tidal Evolution} \label{sec:coupled_evolution} 

The coupled stellar-tidal orbital evolution of binary systems has been extensively studied in the literature for systems ranging from star-star binaries \citep[e.g.][]{Huang1966,Mestel1968,VantVeer1988,Zahn1989,Li1998,Khaliullin2011} to star-planet binaries \cite[e.g.][]{DobbsDixon2004,Barker2009,Lanza2016} to even star-compact object binaries \citep[e.g.][]{Verbunt1981,Repetto2014}.  In particular, the pioneering theoretical study of \citet{Zahn1989} tracked the coupled stellar-tidal evolution of low-mass stellar binaries with a focus on the pre-main sequence evolution and outlined the general qualitative behavior that arises from this coupling.  Although their study mainly focused on orbital circularization during the pre-main sequence, \citet{Zahn1989} also showed that as the binary approached synchronization, the binary orbital period increases as tides transfer rotational angular momentum from the stellar rotations to the orbit.  We reproduce this phenomenon in our simulations (see $\S$~\ref{sec:results}).

Common to many of the aforementioned studies of coupled stellar-tidal evolution is that for tidally locked systems, any angular momentum lost from the star(s) is lost at the expense of the orbit.  For tidally locked synchronous rotators, for example, as magnetic braking slows the stellar rotations, tides speed up the stellar rotations to force the stars' spin periods to be equal to the orbital period.  Tidal speed-up of stellar rotations to maintain synchronization removes angular momentum from the orbit, causing orbital semi-major axis decay and faster stellar rotations.  A similar spin-orbit coupling occurs for stellar radius contraction and expansion for tidally locked stars.  In this section, we derive equations for how the binary semi-major axis, $a$, changes due to both magnetic braking and stellar radius evolution when either one or both stars are in a tidally locked orbit.  We assume constant mass, as mass loss is negligible for low-mass main sequence stars.  These equations provide an additional change in the binary semi-major axis, denoted $\dot{a}_{coupled}$, such that the net change in the binary semi-major axis is $\dot{a}_{net} = \dot{a}_{tides} + \dot{a}_{coupled}$ where the $\dot{a}_{tides}$ term comes from \eqtide Eq.~{\ref{eqn:cpl_dadt_net}}. We explore the dynamical and observational consequences for magnetic braking-driven semi-major axis decay in $\S$~\ref{sec:fiducial_simulation} and $\S$~\ref{sec:obs_acrit}.

\subsubsection{Case 1: One Tidally Locked Star}

In the case of one tidally locked star, the angular momentum of the orbit and star are explicitly coupled as any angular momentum change in the star imparts a change in the orbit as mediated by tidal forces.  In this case, we consider the following quantity
\begin{equation} \label{eqn:ang_mom_one}
J = \mu_{tot} \sqrt{GMa(1-e^2)} + m_1 r_{g,1}^2 R_1^2 \omega + J_{mb},
\end{equation}
where $J$ is the total angular momentum, $J_{mb}$ is the reservoir of angular momentum lost to space via magnetic braking, $\mu_{tot} = m_1 m_2 / (m_1 + m_2)$, $M = m_1 + m_2$ and the rotation rate $\omega$ is set by the star's tidally locked state, e.g. synchronous or a 3:2 spin-orbit-resonance.  This net angular momentum quantity only includes the contributions from both the orbit and the tidally locked star as these are explicitly coupled by tides in this case.  We assume $\dot{J} = 0$ as the total angular momentum is conserved.  By taking the time derivative of Eq.~(\ref{eqn:ang_mom_one}) and rearranging, we obtain
\begin{equation} \label{eqn:tidal_locked_one}
\begin{split}
\dot{a}_{coupled}^{(1)} = \frac{-\dot{J_{mb}} - 2 m_1 r_{g,1}^2 \omega R_1 \dot{R_1} + \frac{\mu^2 G M a e }{J} \dot{e}}
{\frac{\mu^2 G M (1-e^2)}{2J} - \frac{3 \omega}{2a} m_1 r_{g,1}^2 R_1^2}
\end{split}
\end{equation}
for the binary semi-major axis change due to the magnetic braking and stellar radius evolution for one tidally locked rotating star.  Note that $\dot{J_{mb}} > 0$ as this term tracks the amount of angular momentum lost from the system and hence gains the amount lost from stars due to magnetic braking.  Eq.~(\ref{eqn:tidal_locked_one}) is given for the case when the primary star is tidally locked and is trivially altered for the case when the secondary star is tidally locked by exchanging indices. 

Following the lead of previous works \citep[e.g.][]{Verbunt1981,Repetto2014}, we assume magnetic braking and stellar radius evolution do not torque the orbit and hence cannot change $e$, only $a$.  We leave the $\dot{e}$ term in Eq.~(\ref{eqn:tidal_locked_one}) and Eq.~(\ref{eqn:tidal_locked_two}) for completeness.

\subsubsection{Case 2: Two Tidally Locked Stars}

When both stars are tidally locked, the angular momentum of the orbit and both stars is coupled.  Just as in the one-tidally-locked-star case, the total angular momentum of the system 
\begin{equation} \label{eqn:ang_mom_two}
J = \mu_{tot} \sqrt{GMa(1-e^2)} + m_1 r_{g,1}^2 R_1^2 \omega + m_2 r_{g,2}^2 R_2^2 \omega + J_{mb},
\end{equation}
where 1 and 2 denote the primary and secondary star, respectively, and $\dot{J} = 0$.  As before, we take the time derivative of Eq.~(\ref{eqn:ang_mom_two}), rearrange, and obtain
\small
\begin{equation} \label{eqn:tidal_locked_two}
\begin{split}
\dot{a}_{coupled}^{(2)} = \frac{-\dot{J_{mb}} - 2 \omega \left( m_1 r_{g,1}^2 R_1 \dot{R_1} + m_2 r_{g,2}^2 R_2 \dot{R_2} \right) + \frac{\mu^2 G M a e }{J} \dot{e}}
{\frac{\mu^2 G M (1-e^2)}{2J} - \frac{3 \omega}{2a} \left( m_1 r_{g,1}^2 R_1^2 + m_2 r_{g,2}^2 R_2^2 \right)}
\end{split}
\end{equation}
\normalsize
for the binary semi-major axis change due to the magnetic braking and stellar radius evolution.  

\subsubsection{Case 3: Free Rotators}

We include either Eq.~(\ref{eqn:tidal_locked_one}) or Eq.~(\ref{eqn:tidal_locked_two}) in our numerical integration only when either one or both stars become tidally locked in our simulations.  When neither star is tidally locked, the aforementioned mechanism does not apply, and hence we set $\dot{a}_{coupled} = 0$.

\subsection{Energy and Angular Momentum Conservation} \label{sec:conservation}

\begin{figure*}[t]
	\includegraphics[width=\textwidth]{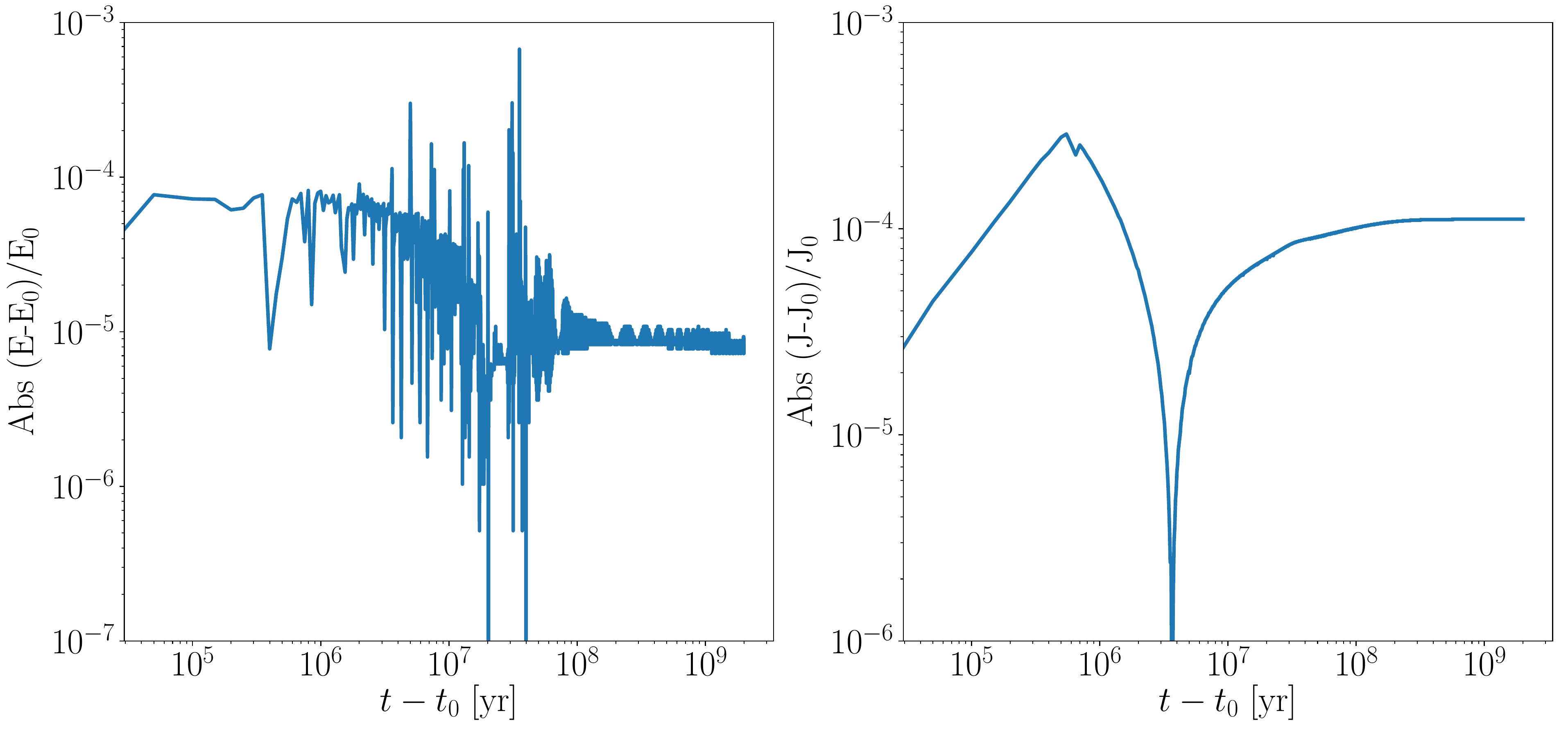}
    \caption{{\it Left Panel:} Absolute relative difference between the total system energy and initial total system energy as a function of time.  {\it Right Panel:} Absolute relative difference of the total system angular momentum and initial total system angular momentum as a function of time.  Both total energy and angular momentum are approximately conserved at the $10^{-4}$ level with high frequency oscillations caused by limitations of Runge-Kutta integrations.}
    \label{fig:conservation}
\end{figure*}

In Fig.~\ref{fig:conservation}, we plot the relative absolute change in both the total system energy and angular momentum as a function of time for our fiducial simulation (see Table~\ref{tab:params}, $\S$~\ref{sec:fiducial_simulation}) to demonstrate that both energy and angular momentum are approximately conserved in our 
\vplanet simulations.  The secular drifts in total angular momentum stem from our usage of the CPL tidal model that extends to only second order in $e$ and hence cannot not exactly conserve angular momentum.  We conclude our methodology satisfies the conservation laws to sufficient accuracy.

\subsection{N-body Simulations of Circumbinary Planetary Systems} \label{sec:nbody_methods}

Here, we complement our \vplanet simulations of coupled stellar-tidal evolution with several suites of N-body simulations that probe the dynamical and observational impact of single and two-planet circumbinary planetary systems in which the inner-most planet falls within the $a_{crit}$.  We simulate such systems using the N-body code \texttt{REBOUND} \citep{Rein2012}.  We use \texttt{REBOUND}'s high-order, adaptive timestepping \texttt{IAS15} integration scheme to integrate the gravitational forces in our simulations as it is accurate and flexible enough to handle close encounters and scattering events while conserving energy to a high precision \citep{Rein2015}.  We outline the simulation setup and our choice of initial conditions in $\S$~\ref{sec:nbody_initial_conditions}.

We show that due to coupled stellar-tidal evolution, a CBP that initially resides near the dynamical stability limit can fall within the $a_{crit}$ and go unstable, potentially disrupting the entire circumbinary planetary system.  For single-planet circumbinary systems, the planet will likely be ejected, potentially leaving no observational signature that the binary ever hosted a planet.  Multi-planet circumbinary systems in which the inner planet falls within $a_{crit}$, however, can potentially undergo richer dynamical evolution in which binary-planet scattering and planet-planet scattering can reshape the system resulting in multiple planetary ejections, but also leave one or more planets bound to the stars.  Several previous studies have examined the impact of binary-planet and/or planet-planet scattering in circumbinary systems \citep[e.g.][]{Kratter2014,Smullen2016,Sutherland2016,Gong2017,Gong2017b}.  No study yet has examined the case we consider in this paper in which the inner planet of a two-planet circumbinary planetary system initially resides interior to the dynamical stability limit.  This calculation is important as it not only allows us to estimate how many planets are actually ejected from these systems but also to estimate how ejection and scattering events can impact the detectability of any surviving CBPs.

Note that in our simulations, we do not model planet-planet or star-planet collisions for computational speed and simplicity.  This approach is justified as previous studies of planet-planet scattering and instabilities in circumbinary planetary systems all find that planetary ejection occurs far more often that collisions when a planet is lost from the system \citep[see][]{Smullen2016,Sutherland2016}.  Also note we do not consider the coupled stellar-tidal-N-body evolution.


\section{Simulations} \label{sec:simulations}

Here we outline the details of both our \vplanet and \texttt{REBOUND} N-body simulations and justify our assumed initial conditions.

\subsection{Coupled Stellar-Tidal Initial Conditions} \label{sec:initial_conditions}

We probe how model parameters and the underlying assumptions impact the ability for coupled stellar-tidal evolution to destabilize CBPs by running several sets of simulations varying just one parameter at a time and several suites of many simulations varying multiple parameters over the adopted ranges.  Using \vplanet, we simulated coupled stellar-tidal evolution by simultaneously integrating the equations presented in $\S$~\ref{sec:stellar_evolution}-\ref{sec:coupled_evolution} for 2 Gyr for each simulation as described in $\S$~\ref{sec:methods}.  In Table~\ref{tab:params}, we give the default values and range for our simulation initial conditions.  Below, we justify our choice of parameter ranges and fiducial values and discuss the qualitative results of a simulation initialized with the fiducial values to elucidate how coupled stellar-tidal evolution drives changes in $a_{crit}$.

\begin{deluxetable}{lcc}
\tabletypesize{\small}
\tablecaption{Parameter Ranges \label{tab:params}}
\tablewidth{0pt}
\tablehead{
\colhead{Parameter} & \colhead{Range} & \colhead{Fiducial}
}
\startdata
$M_\star$ [$M_{\odot}$] & $0.5 - 1$ & $1$ \\  
$e$ & 0.0 - 0.2 & 0.15 \\
$P_{bin,init}$ [d] & $3 - 7.5$ & 4 \\
$P_{rot,init}$ [d] & $0.1 - 2$ & 0.5 \\
$Q$ & $10^5 - 10^7$ & $10^6$ \\
$r_g$ & $0.15 - 0.45$ & $0.27$
\enddata \vspace*{0.1in}
\end{deluxetable}

\subsubsection{Stellar Tidal Qs} \label{sec:tidal_q}

Binary stars' tidal Qs strongly determine the extent to which a short-period binary tidally evolves.  Currently, the value of Q for low-mass stars is uncertain.  From observations of tidally circularized binaries in clusters, tidal Qs for sun-like stars have been estimated to be of order $Q{\sim}10^6$ \citep[e.g.][]{Meibom2005}.  Observations of orbital decay of hot Jupiters find tidal Qs for the host stars are of order ${\sim}10^5-10^6$ \citep{Jackson2009,Essick2016,Patra2017,Wilkins2017}.  Given these observations, we adopt $Q=10^6$ as our fiducial value.  Detailed studies of tidal dissipation in sun-like stars by \citet{Ogilvie2007}, however, have shown that a star's tidal Q has complicated dependencies on the viscous and hydrodynamical processes operating within the star and can strongly vary depending on the star's spin frequency and the orbital frequency of its companion.  Additionally, \citet{Barker2009} find that the tidal Q likely varies even among stars in the same spectral class.  In the case of pre-main sequence stars, \citet{Bolmont2016} find that efficient tidal dissipation yields tidal Qs of order $10^4-10^6$.  Since our simple model uses a constant tidal Q and cannot capture more complex tidal Q evolution, in $\S$~\ref{sec:var_Q} we vary the tidal Q amongst our simulations by two orders of magnitude to probe our model's sensitivity to various tidal Qs.

\subsubsection{Stellar Rotations} \label{sec:stellar_rotations}

The STEEP process requires that short-period binary stars form with rotation periods ($P_{rot}$) shorter than the orbital period.  For short stellar rotational periods, $P_{rot} \lsim 1$ day, the rotational angular momentum can be of order the orbital angular momentum, allowing for a significant increase in $a_{crit}$ via tidal transfer of angular momentum.  Here, we review observations of young single stars and binaries in open clusters to justify our stellar rotation assumptions.

In a study of about 250 stars in the ${\sim}1$ Myr old Orion OBIc/d association, \citet{Stassun1999} find a flat distribution for $P_{rot} > 0.5$ days.  Interestingly, \citet{Stassun1999} find several stars rotating near the break-up velocity, a rotational period of $P_{rot}{\sim} 0.25$ days, indicating that young stars can indeed be very rapid rotators.  Given this finding, we set the lower limit for initial $P_{rot}$ in our study to be $0.1-0.2$ days.  \citet{Rebull2006} examined the $P_{rot}$ of about 900 stars in the roughly 1 Myr old Orion Molecular Cloud, finding most stars with $1 < P_{rot} < 10$ days, and a minority with $P_{rot} \lsim 1$ day.  A study of $P_{rot}$ of young weak T Tauri star candidates in the Orion star forming region by \citet{Marilli2007} found a roughly flat distribution with a peak near $P_{rot} = 1.5$ days, similar to the aforementioned findings.  Studies of older open clusters with ages ${\sim}100-200$ Myr consistently find stellar rotation distributions with $P_{rot}$ often a low as $P_{rot} = 0.5$ days and many stars with $P_{rot} = \lsim 1$ day, although there is an appreciable spread in $P_{rot}$ with some reaching up to $P_{rot} {\sim} 10$ days \citep[see][]{Marilli2007,Meibom2009,Meibom2011}.  Young stars can readily have $P_{rot} \lsim 1$ day.

For the $P_{rot}$ of binary stars, \citet{Meibom2007} observed the ${\sim}150$ Myr old open cluster M35 and measured the $P_{rot}$ distributions of primary stars of close binaries and single stars.  \citet{Meibom2007} found that the primary stars in binaries tended to have shorter $P_{rot}$ than single stars with statistically significant differences in the means and medians of the two $P_{rot}$ distributions of at least the $99.9\%$ level after controlling for tidal effects.  \citet{Stauffer2016} derive a similar result from observations of $P_{rot}$ in young, low-mass binaries in the Pleiades.  We conclude that a significant number of young binary stars form with $P_{rot} \lsim 1$ days. 

\subsubsection{Stellar Radius of Gyration} \label{sec:r_g}

Measuring the stellar $r_g$ is, in general, quite a difficult task so here we rely on theoretical stellar evolution models to inform our choices.  From the \citet{Baraffe2015} stellar evolution models for sun-like stars, $r_g{\approx}0.45$ on the pre-main sequence, decreasing to $r_g{\approx}0.27$, our fiducial value, on the main sequence.  Our model utilizes a constant $r_g$, so we vary it in a series of simulations to gauge how strongly it impacts the evolution of $a_{crit}$ over time in $\S$~\ref{sec:var_rg}.  We consider our fiducial value of $r_g = 0.27$ to be a conservative estimate as we show in $\S$~\ref{sec:results} most of the $a_{crit}$ evolution occurs while the stars reside on the pre-main sequence.

\subsection{N-body Simulations}

Each N-body simulation is comprised of two segments.  First, we run long-term integration probing the dynamical stability of a one- or two-planet circumbinary system comprised of planets ``b" and ``c", when applicable, to determine if any planets are ejected from the system.  Once that run finishes, we run a short-term integration initialized with the final state of the long-term dynamical stability integration to perform a series of mock transit observations that explore how dynamical instabilities in such systems impact the detectability of the remaining planets.  Splitting a given simulation into two segments allows us to both characterize the dynamics of circumbinary systems in which the inner planet falls within $a_{crit}$ and to probe the resulting observational consequences.  We outline the procedures for each part of the simulation below.  We summarize and examine the results of these simulations in $\S$~\ref{sec:nbody_results}.

\subsubsection{N-body Simulations Initial Conditions} \label{sec:nbody_initial_conditions}

For each simulation, we initialize the binary with two $1$ M$_{\odot}$ stars with an orbital period of $7$ days and $e$ randomly uniformly sampled from $[0,0.2]$ orbiting in the $x-y$ plane.  This choice simplifies the system geometry such that the binary is always in a transiting configuration ($i_{bin} = 90^{\circ}$).  The binary longitude of the ascending node ($\Omega$), argument of pericenter ($\omega$), and mean anomaly ($M$) are all randomly uniformly sampled from $[0,2\pi)$.

With the binary parameters set, b's semi-major axis is randomly uniformly sampled from $[0.94,1]$ times $a_{crit}$. i.e. just within the $a_{crit}$ (recall that the errors on Eq.~(\ref{eqn:crit_semi}) are at worst $6\%$ \citep{Holman1999}).  In multi-planet simulations, planet c's semi-major axis is constructed such that it is randomly uniformly sampled from $[5,10]$ mutual Hill radii from b, where a mutual Hill radius is
\begin{equation} \label{eqn:mutual_hill_radius}
R_{hill, mutual} = \left( \frac{m_b + m_c}{3M_{bin}} \right)^{\frac{1}{3}} \frac{a_b + a_c}{2},
\end{equation}
where $m_b$ and $m_c$ are the masses of planet b and c, respectively, $M_{bin}$ is the total mass of the binary, and $a_b$ and $a_c$ are the semi-major axes of planets b and c, respectively \citep{Chambers1996}.  CBP eccentricties are randomly uniformly sampled from $[0,0.1]$ and $\Omega$, $\omega$ and $M$ are all randomly uniformly sampled from $[0,2\pi)$.  We summarize the randomized orbital elements for both the binary and the planets in Table~\ref{tab:nbody_params}.  All CBP orbital elements are initialized in Jacobi coordinates.

In single planet simulations, we examined three cases varying the mass of planet b, while in multi-planet simulations, we examined six cases in which we varied both the mass of planets b and c and their initial inclination relative to the plane of the binary to examine their impact on our results.  For the planet masses, we consider three cases: Neptune-, Saturn-, and Jupiter-mass CBPs as these roughly span the observed masses for transiting CBPs.  Both planet masses are randomly sampled from the same normal distributions with mean $m_p$ and standard deviation $0.1m_p$, where $m_p$ is equal to the mass of Neptune, Saturn, or Jupiter depending on the simulation suite.  We chose to sample both planet masses from the same distribution for a given simulation to keep the planet mass ratio $m_b/m_c {\sim} 1$, which simplifies our results because varying CBP mass ratios can strongly impact the scattering process in such systems \citep{Gong2017}.  For the planet inclinations relative to the binary, we randomly uniformly sample from $[0^{\circ},1^{\circ}$] (the ``low inclination" case) or $[0^{\circ},3^{\circ}$] (the ``high inclination" case).  Both of these initial inclination distributions are broadly consistent with the observed trend of transiting circumbinary exoplanets to be nearly coplanar with their host binary \citep[e.g.][]{Li2016}, but we stress that the true inclination distribution of transiting circumbinary exoplanets is unknown.  We adopt these two simple distributions given our ignorance of the true underlying distribution and since the inclination distribution can significantly impact the observed transitablility of circumbinary exoplanets \citep[see][]{Armstrong2014,Martin2015a}.  We run 1,000 simulations for each of the aforementioned cases for a total of $9,000$ N-body simulations.

\begin{deluxetable}{lc}
\tabletypesize{\small}
\tablecaption{N-body Simulation Initial Condition Ranges \label{tab:nbody_params}}
\tablewidth{0pt}
\tablehead{
\colhead{Parameter} & \colhead{Distribution}
}
\startdata
$e_{bin}$ & $U(0,0.2)$ \\  
$e_{b,c}$ & $U(0, 0.1)$ \\
$a_{b}$ [AU] & $U(0.94, 1.0) \times a_{crit}$ \\
$n R_{hill, mutual}$ & $U(5, 10)$ \\
$\Omega_{bin,b,c}$ & $U(0, 2\pi)$ \\
$\omega_{bin,b,c}$ & $U(0, 2\pi)$ \\
$M_{bin,b,c}$ & $U(0, 2\pi)$ 
\enddata \vspace*{0.1in}
\end{deluxetable}

\subsubsection{Dynamical Stability Integration}

In the first part of a simulation, we integrate a one or two-planet circumbinary system for $10^5$ binary orbital periods.  This timescale, about an order of magnitude longer than the CBP dynamical stability simulations of \citet{Holman1999} in terms of binary orbital periods, and is sufficiently long for the majority of dynamically unstable systems to go unstable.  We classify a system as unstable when one or both of the planets is ejected from the system.  A planet is considered ejected when its semi-major axis exceeds $50$ AU from the system barycenter, a distance that is over an order of magnitude larger than any of the CBP's initial semi-major axes.  When a planet gets ejected, it is removed from the N-body simulation.  At the end of the integration, we record the final architecture of the simulation, namely the remaining planet's Cartesian positions and orbital elements relative to the binary barycenter.  

\subsubsection{Mock Transit Observation Integration}

In the second part of a simulation, we integrate the remaining bodies in the system for $4$ years, the approximate lifetime of the \textit{Kepler} mission, and perform mock transit observations to estimate if any remaining CBPs transit and, if so, how frequently they transit.  The geometry of our simulations is set up such that the binary orbits in the $x-y$ plane and the observer looks down the $+x$ axis towards the origin.  In this simplified configuration, the binary is always in a transiting configuration and almost all of the CBPs are initially in a transiting configuration.  A more realistic treatment would allow the binary to have an arbitrary inclination on the sky with respect to the observer and include more physically-motivated binary and CBP orbital parameter priors.  For our purposes, however, this simple case permits a first order analysis of the dynamics of these unstable systems and their observational consequences. We leave a more robust treatment for future work and refer the reader to \citet{Martin2014} for a detailed examination of the detectability of CBPs orbiting non-transiting binaries.

To perform mock transit observations, every ${\sim}2$ simulation minutes we record if any of the remaining planets are transiting either of the two host stars.  A planet is transiting if
\begin{equation} \label{eqn:transit}
d < r_{planet} + r_{star} \text{ and } x_{planet} > x_{star}
\end{equation}
where $r_{i}$ is the radius of the $i^{th}$ body, $d$ is the projected distance between the centers of mass of the planet and star under consideration, and $x_{i}$ is the $x$ Cartesian coordinate of the $i^{th}$ body.  We require $x_{planet} > x_{star}$ since in our simplified geometry, the observer lies along the $+x$ axis and looks towards the origin.  This frequent sampling over the course of the \textit{Kepler} timescale integration not only checks if any remaining planet transits, but also facilitates the calculation of the fraction of time transiting either or both of the binary stars (henceforth referred to as FTT).  Given the inherent difficulty in detecting transiting CBPs \citep[e.g.][]{Welsh2014,Winn2015}, FTT is a useful quantity as CBPs that have larger FTTs spend more time transiting and should be more detectable.  We discuss the results of this analysis in $\S$~\ref{sec:nbody_results}.

\subsection{Tying It All Together}

In summary, we first run an expansive set of \vplanet simulations to examine how coupled-stellar tidal evolution affects $a_{crit}$.  Our simulations not only probe how this evolution depends on parameters such as the stellar tidal Q, but also examine thousands of different initial states for diverse binary systems.  These simulations reveal how the STEEP process forces CBPs within $a_{crit}$.  We present the results of these simulations in $\S$~\ref{sec:results}.  To complete the theoretical picture given by the STEEP process, we then run an ensemble of N-body simulations with \texttt{REBOUND} to characterize how CBP systems evolve when one planet is interior to the stability limit.  Finally, we follow up these simulations with mock transit observations to gain a crude understanding of how the STEEP process impacts the observability of any surviving CBPs.  We present the results of these simulations in $\S$~\ref{sec:nbody_results}.


\section{Results: Coupled Stellar-Tidal Evolution Simulations} \label{sec:results}

Here we present the results of simulations of coupled stellar-tidal evolution of binary star systems.  In $\S$~\ref{sec:fiducial_simulation} through $\S$~\ref{sec:optimistic} we simulate coupled stellar-tidal evolution for binary star systems using \vplanet to quantify how the initial binary orbit, the initial stellar rotations, and the details of the tidal interactions impact the evolution of $a_{crit,init}$.

\subsection{Fiducial Simulation} \label{sec:fiducial_simulation}

\begin{figure}[h]
	\includegraphics[scale=0.5]{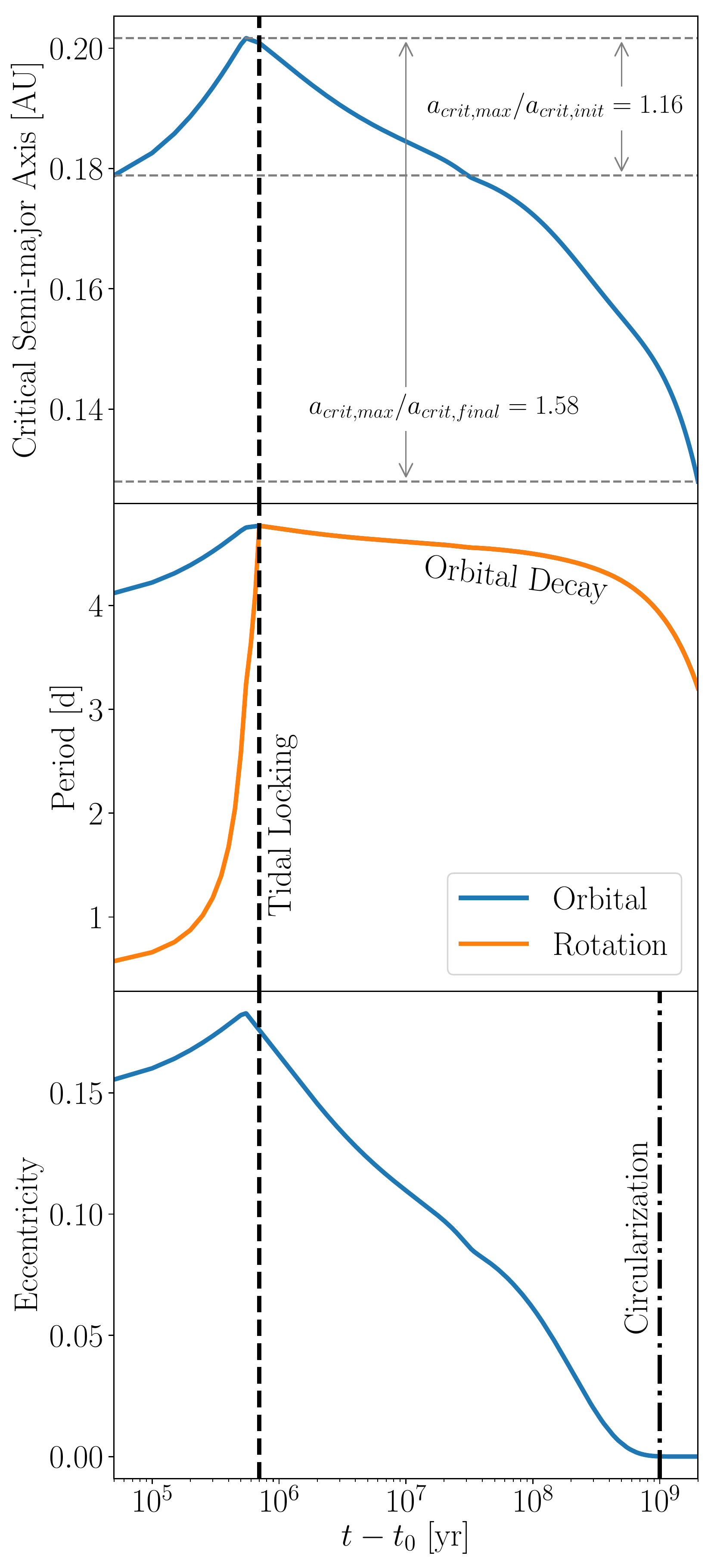}
   \caption{Evolution of a binary system using our default parameters from Table~\ref{tab:params}.  {\it Top:} $a_{crit}$ vs. time.  The grey arrows demarcate both the ratio of the maximum critical semi-major axis to the initial and final values, $a_{crit,max}/a_{crit,init} = 1.16$ and $a_{crit,max}/a_{crit,final} = 1.58$ for this system, respectively. {\it Middle:} Orbital and stellar rotation period vs. time.  The black dashed line indicates that the binary tidally locks after about 1 Myr.  {\it Bottom:} $e$ vs. time.  The dot-dashed line indicates when the binary circularizes at about 1 Gyr, well after tidally locking.}
    \label{fig:example}
\end{figure}

To demonstrate how binary star systems evolve due to coupled stellar-tidal physics, we present the full evolution of a $1$ M$_{\odot}$$-1$ M$_{\odot}$ binary system using the fiducial parameter values given in Table~\ref{tab:params} for the initial conditions (compare to Fig.~\ref{fig:eqtide_example}).  We present the results of this simulation in terms of $a_{crit}$, the orbital and rotation periods, and the $e$ in Fig.~\ref{fig:example}, annotated with key simulation results and evolutionary regimes.

Initially,  the rotation rate, $\omega$, slows as tides transfer rotational angular momentum into the orbit causing the orbital period to grow.  During this time, stellar contraction supplies additional rotational angular momentum, slowing the $\omega$ decay, while magnetic braking removes some rotational angular momentum from the system entirely.  The system tidally locks after about 1 Myr as tides efficiently transport stellar rotational angular momentum into the orbit.

The early growth in the binary orbital period drives most of the $a_{crit}$ growth with the small increases in $e$ providing the rest.  $a_{crit}$ reaches its largest value just before tidal locking occurs due to the slight $e$ growth.  For $e=0$, the binary would reach the peak $a_{crit}$ precisely once the binary tidally locks.  In the top panel of Fig.~\ref{fig:example}, we show in grey the difference between the maximum and initial critical semi-major axis, $a_{crit,max}/a_{crit,init} = 1.16$, a value likely large enough to destabilize some CBPs, see Fig.~\ref{fig:observed_acrit}.  

In all our simulations, we find that a binary reaches the maximum semi-major axis at about the time it becomes tidally locked.  This occurs for two reasons.  First, once both stars are tidally locked and synchronized, tides have already transferred as much angular momentum from the stellar rotations into the orbit as possible - any remaining tidal coupling will work to maintain the tidal locking and will not further expand the binary orbit.  Second, for these systems, we find that tidal locking occurs well in advance of orbital circularization via tides.  Larger non-zero eccentricities maintain larger $a_{crit}$ according to Eq.~(\ref{eqn:crit_semi}). 

After tidal locking, the binary rotation and orbital periods perfectly mirror each other as tides keep the binary synchronous (see $\S$~\ref{sec:tidal_evolution} and $\S$~\ref{sec:32} for higher-order spin-orbit resonances at larger $e$).  In this regime, $e$ decreases until the orbit is circularized after about 1 Gyr, well after the stars are tidally locked, further decreasing $a_{crit}$.

Once the system is tidally locked, magnetic braking cannot slow stellar rotations, so instead it removes angular momentum from the orbit causing the orbital period and $a_{crit}$ to decay dramatically with the orbital period dropping by almost 1 day per Gyr.  This substantial orbital decay causes $a_{crit}$ to drop by about a factor of 1.6 relative to its maximum value.  The combination of orbital circularization and magnetic braking for tidally locked binaries causes significant observational consequences for short-period binaries: the orbit, and hence $a_{crit}$, observed today is likely much different than what it was in the past.  For example if this system was observed at the end of the simulation, we might expect to find CBPs near $a_{crit}$ based on the {\it Kepler} CBP discoveries.  Coupled stellar-tidal evolution could have destabilized CBPs near $a_{crit}$ early on in the system's lifetime such that there would be no CBPs to detect.  $a_{crit}$ decay implies that the dynamical instability region around the binary was much larger in the past such that any surviving CBPs would necessarily have to be located at larger $a_{CBP}$ relative to the central binary's $a$ and hence be harder to detect.  We examine this effect further in $\S$~\ref{sec:obs_acrit} with a particular focus on the impact of the initial binary orbital period and the details of the magnetic braking physics.

\subsection{Varying Tidal Q} \label{sec:var_Q}

\begin{figure*}[t]
	\includegraphics[width=\textwidth]{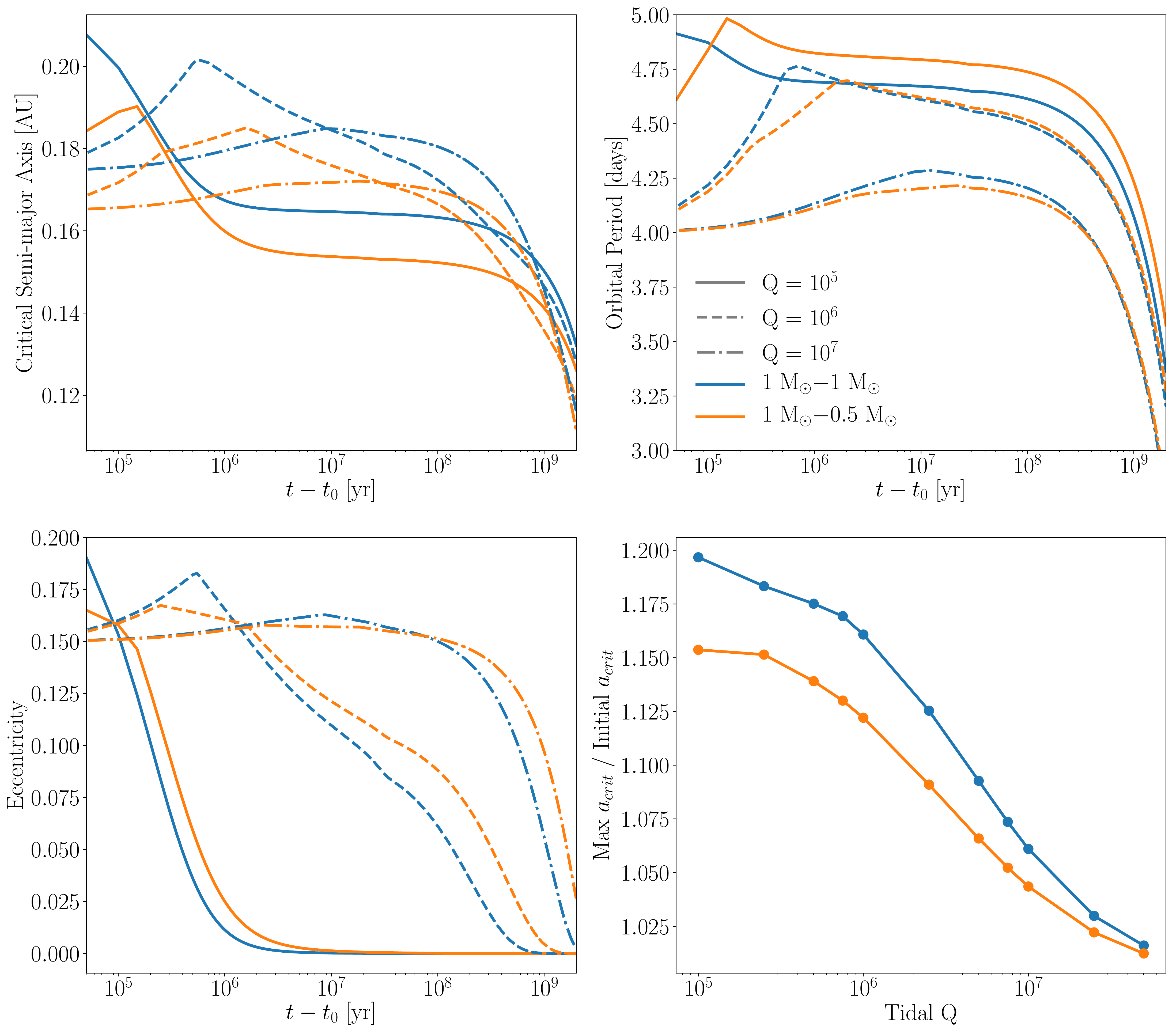}
    \caption{The role of tidal dissipation in binary star evolution. {\it Top Left:} $a_{crit}$ vs. time.  {\it Top Right:} Binary orbital period vs. time. {\it Bottom Left:} $e$ vs. time. {\it Bottom Right:} $a_{crit,max}$/$a_{crit,init}$ vs. tidal Q.  The blue and orange lines correspond to a $1$ M$_{\odot}$$-1$ M$_{\odot}$ and a $1$ M$_{\odot}$$-0.5$ M$_{\odot}$ binary, respectively.  The solid, dashed, and dot-dashed lines correspond to tidal Qs of $10^5$, $10^6$, and $10^7$, respectively.  Smaller tidal Qs drive large $a_{crit}$ expansion.}
    \label{fig:var_Q}
\end{figure*}

The value of the stellar tidal Qs is of primary importance to the STEEP process as it controls the timescale of tidal evolution and the rate of angular momentum transfer from stellar rotations to the orbit.  In Fig.~\ref{fig:var_Q}, we present the evolution of a $1$ M$_{\odot}$$-1$ M$_{\odot}$ and a $1$ M$_{\odot}$$-0.5$ M$_{\odot}$ binary with stellar tidal Qs ranging from $10^5$-$10^7$.  For simplicity, we set both stars' Q to the same value.  

For each simulation, the binary orbital period increases by upwards of one day for smaller tidal Qs as angular momentum is tidally transferred to the orbit.  Stellar contraction during the pre-main sequence phase speeds up the stellar rotations, providing additional angular momentum for transfer into the orbit.  The increasing orbital period is accompanied by a modest increase in $e$ from 0.15 to upwards of 0.18.  The orbital period increases until the system tidally locks at which point the tidal transfer of rotational angular momentum to the orbit is complete.  Tidal locking occurs at around $10^5$ years for $Q=10^5$ and 1 Myr and 10 Myr for $Q=10^6$ and $Q=10^7$, respectively.  The increasing orbital period primarily drives the increase in $a_{crit}$, which peaks approximately at the same time as the orbital period.  Once the system tidally locks, the orbital period decays as magnetic braking siphons angular momentum from the orbit.  As seen in Fig.~\ref{fig:var_Q}, magnetic braking drains a large amount of angular momentum from the orbit, often causing the orbital period to decrease by a little over a day in 100 Myr, a decrease of $25\%$.

We find that binary stars with lower tidal Qs tend to reach larger $a_{crit,max}$/$a_{crit,init}$.  Lower tidal Qs lead to faster tidal evolution since the equations in our CPL model all have a $1/Q$ dependence (see $\S$~\ref{sec:tidal_evolution}).  When the tidal evolution proceeds more quickly, there is less time for magnetic braking to siphon rotational angular momentum away from the stars, allowing for tides to transport a larger amount of the rotational angular momentum into the orbit, increasing the period and $a_{crit,max}$/$a_{crit,init}$.  Binary stars with $Q \lsim 10^6$ readily reach $a_{crit,max}$/$a_{crit,init} \gsim 1.1$, a value large enough to destabilize some CBPs that form near the dynamical stability limit.

Orbital circularization via tides proceeds more quickly for binary stars with lower stellar tidal Q due to the $1/Q$ scaling in the CPL model equations (see $\S$~\ref{sec:tidal_evolution}).  The orbits of binary stars with $Q = 10^5$ circularize in around $10$ Myr while those with $Q = 10^7$ can take longer than $2$ Gyr to full circularize.  In general, the tidal circularization timescale also depends on the binary orbital period as tighter binaries will circularize more rapidly.  Binary star orbits with orbital periods $\gsim 10$ days are not likely to circularize \citep[e.g.][]{Zahn1989,Meibom2005,Raghavan2010,Lurie2017}.

The unequal-mass binaries tend to have lower $a_{crit,max}$/$a_{crit,init}$ than their equal-mass counter parts because the lower-mass star tidally locks earlier than the more massive primary.  At this point, magnetic braking cannot slow down the lower-mass star's rotation rate as tidal locking fixes it to the mean motion, i.e. magnetic braking siphons angular momentum from the orbit, slowing the growth of the binary period (see $\S$~\ref{sec:coupled_evolution}).

\subsection{Varying Radius of Gyration} \label{sec:var_rg}

\begin{figure*}[t]
	\includegraphics[width=\textwidth]{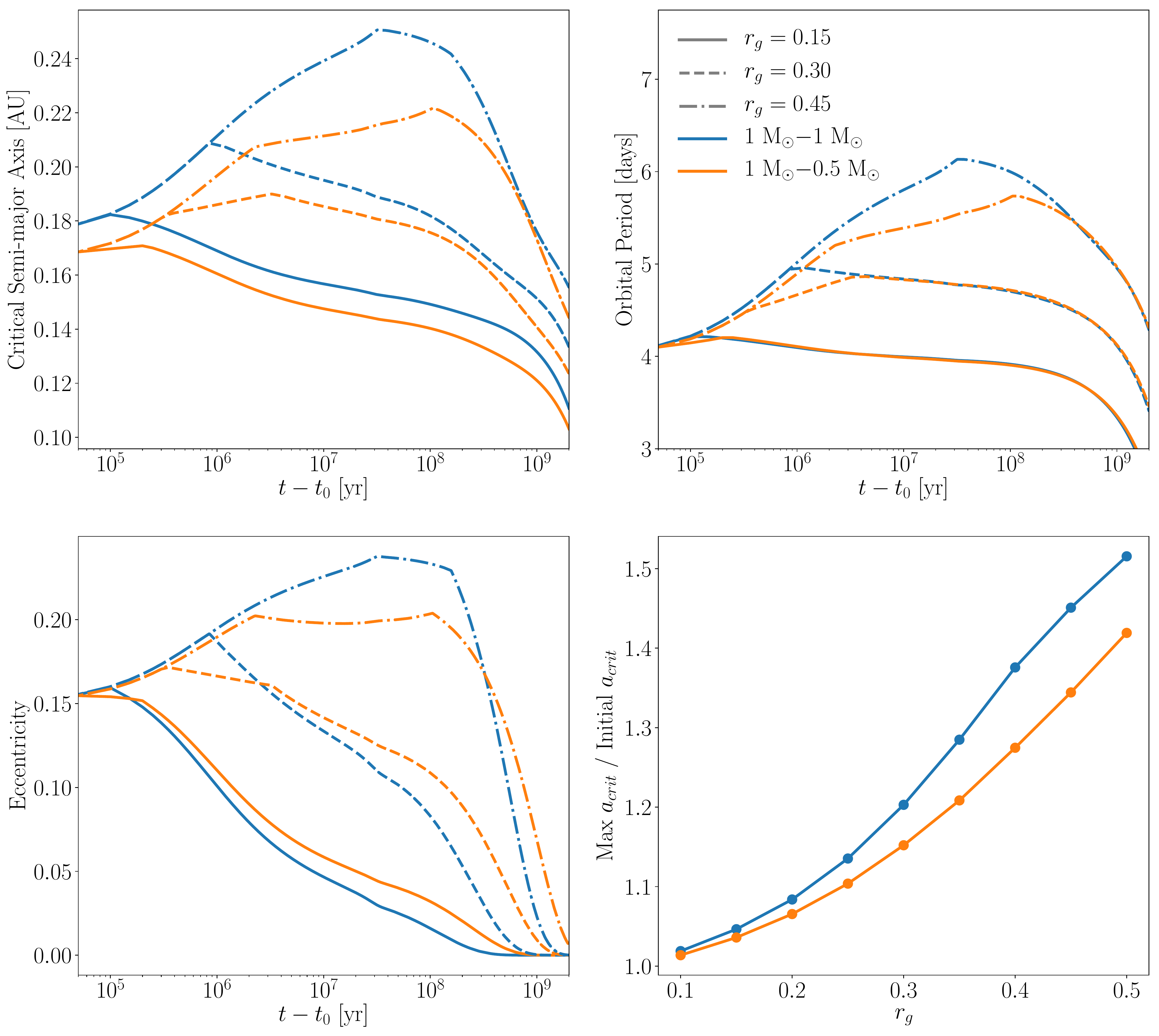}
    \caption{The role of the radius of gyration in binary star evolution. {\it Top Left:} $a_{crit}$ vs. time.  {\it Top Right:} Binary orbital period vs. time. {\it Bottom Left:} $e$ vs. time. {\it Bottom Right:} $a_{crit,max}$/$a_{crit,init}$ vs. $r_g$.  The blue and orange lines correspond to a $1$ M$_{\odot}$$-1$ M$_{\odot}$and a $1$ M$_{\odot}$$-0.5$ M$_{\odot}$ binary, respectively.  The solid, dashed, and dot-dashed lines correspond to an $r_g$ of $0.15$, $0.3$, and $0.45$, respectively.  Larger $r_g$ lead to increased $a_{crit}$ growth.}
    \label{fig:var_rg}
\end{figure*}

The parameter $r_g$ strongly influences how this transfer occurs due to both tidal evolution, see Eq.~(\ref{eqn:cpl_dwdt}), and stellar evolution for magnetic braking (Eq.~(\ref{eqn:stellar_rot_rate_dt})).  In this subsection, we present the full evolution of simulations in which we vary $r_g$ holding all other parameters constant, see Fig.~\ref{fig:var_rg}.

As shown in Fig.~\ref{fig:var_rg}, simulations with large $r_g$ can reach orbital periods over 2 days larger than the initial value primarily due to the stars' rotational angular momentum scaling as $r_g^2$.  With more rotational angular momentum to transfer to the orbit, tidal locking occurs later, at around 10-100 Myr for stars with $r_g=0.45$ as compared to around $10^5$ years for stars with $r_g=0.15$.  Interestingly, all systems in this set of simulations circularize after about 1 Gyr.  Systems with larger $r_g$ circularize much more rapidly once $e$ begins to decrease.  As before, the orbital period growth leads to $a_{crit}$ growth with a peak when the binary tidally locks.  Once tidal locking occurs for both stars, magnetic braking saps angular momentum from the orbit, rapidly decreasing the orbital period by over 1 day over 100 Myr.

Systems with unequal mass binaries tend to reach lower $a_{crit,max}$/$a_{crit,init}$ at a given $r_g$ than their equal mass counterparts because the lower mass star tidally locks more quickly than the more massive primary.  Once tidally locked, the secondary star cannot spin down via magnetic braking so angular momentum instead comes from the orbit, reducing the amount by which the orbital period can grow before the primary star tidally locks.

In general, we find that stars with larger $r_g$ tend to reach larger $a_{crit,max}$/$a_{crit,init}$.  At larger values of $r_g$ for a given rotation rate, a star possesses more rotational angular momentum available for transfer into the orbit via tides allowing for larger $a_{crit,max}$/$a_{crit,init}$.  For binary stars with $r_g \gsim 0.27$, the binaries attain $a_{crit,max}$/$a_{crit,init} \gsim 1.1$, large enough to destabilize and eject some CBPs.  

For sun-like stars, the pre-main sequence lasts for ${\sim} 50$ Myr and up to 1 Gyr for lower-mass late M-dwarfs.  As shown above, a significant portion of  $a_{crit}$ growth occurs while the stars reside on the pre-main sequence as radius contraction provides a substantial angular momentum reservoir for transfer into the orbit via tides.  On the pre-main sequence, \citet{Baraffe2015} predicts that low mass stars have $r_{g}{\sim}0.45$.  Therefore during the pre-main sequence when $a_{crit}$ tends to increase the most, the stellar radius of gyration is large allowing $a_{crit}$ to grow significantly as seen in Fig.~\ref{fig:var_rg} yielding ratios upwards of $a_{crit,max}$/$a_{crit,init} {\sim} 1.4$.  Once the stars reach the main sequence, $r_{g}$ drops to ${\sim}0.27$ slowing the $a_{crit}$ evolution.  Given that we adopted $r_{g}{\sim}0.27$ as our fiducial value when it is likely much larger during the major period of $a_{crit}$ growth, we consider our $a_{crit,max}$/$a_{crit,init}$ estimates to be conservative lower limits. 

\subsection{Varying Magnetic Braking Law} \label{sec:var_mb}

\begin{figure*}[t]
	\includegraphics[width=\textwidth]{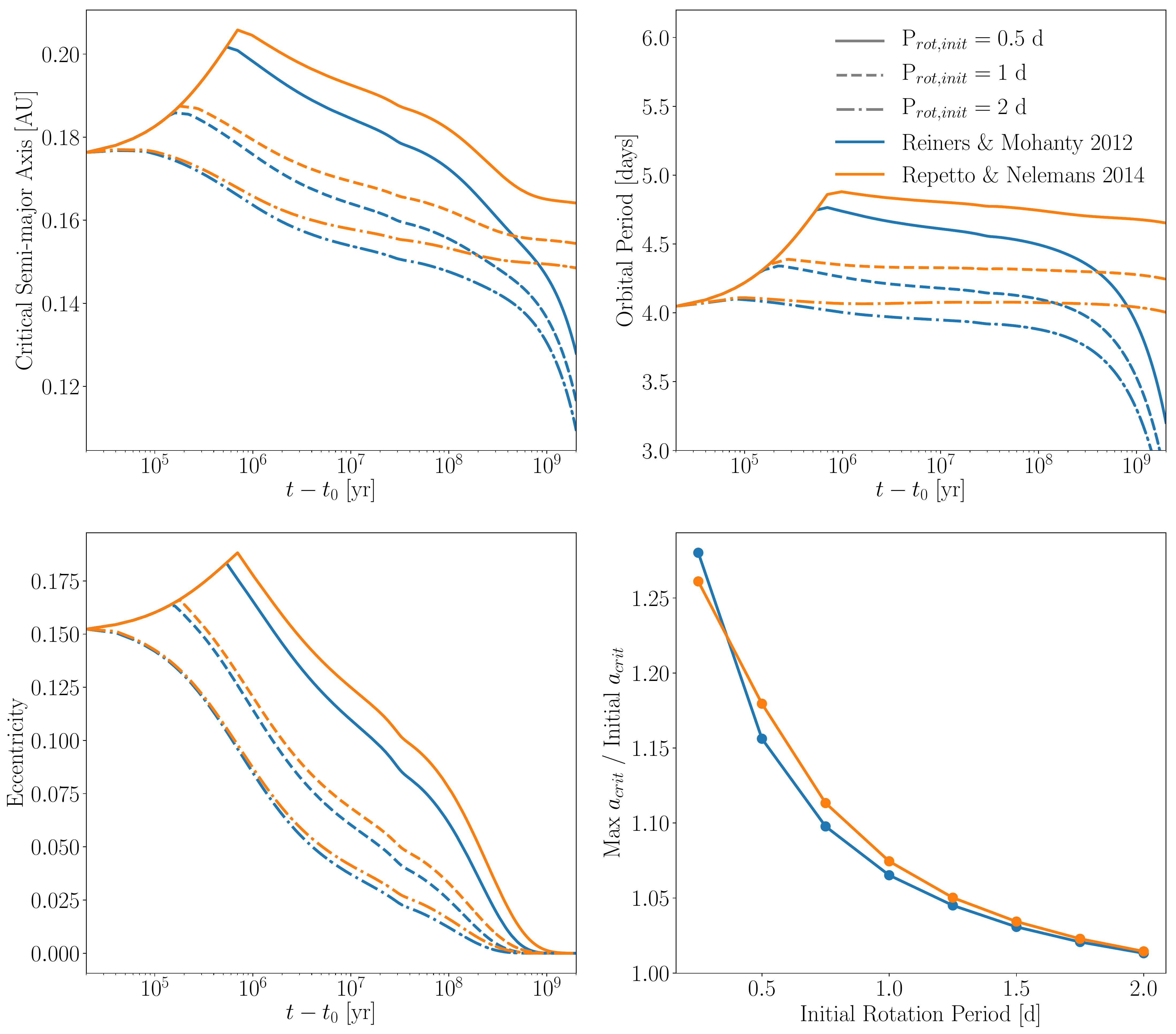}
   \caption{The role of the magnetic braking law and initial stellar $P_{rot}$ in binary star evolution.{\it Top Left:} $a_{crit}$ vs. time.  {\it Top Right:} Binary orbital period vs. time. {\it Bottom Left:} $e$ vs. time. {\it Bottom Right:} $a_{crit,max}$/$a_{crit,init}$ vs. initial stellar $P_{rot}$.  The blue and orange lines correspond to simulations that use the stellar magnetic braking relation from \citet{Reiners2012} and \citet{Repetto2014}, respectively.  The solid, dashed, and dot-dashed lines correspond to a binary with stellar initial spin periods of $0.5$ d, $1$ d, and $2$ d, respectively.  We find that the \citet{Repetto2014} relation leads to larger $a_{crit,max}$/$a_{crit,init}$ and stars with shorter initial spin periods lead to larger $a_{crit,max}$/$a_{crit,init}$ as well.}
    \label{fig:var_mb}
\end{figure*}

Magnetic braking removes angular momentum from stars, slowing their rotation rates.  For the short-period binaries considered here, magnetic braking depletes the stellar rotational angular momentum reservoir that is available for tidal transfer in the orbit before tidal locking, reducing $a_{crit,max}$/$a_{crit,init}$.  To probe the sensitivity of our results to our choice of magnetic braking law, we test run identical sets of simulations using the \citet{Reiners2012} and \citet{Repetto2014} magnetic braking models (see $\S$~\ref{sec:stellar_evolution}) for binaries stars with various initial $P_{rot}$.  Again for simplicity, in each simulation both stars start with the same given $P_{rot}$.  The results of these simulations are shown in Fig.~\ref{fig:var_mb}.  Note that we examine the long-term evolutionary differences between these two magnetic braking models in $\S$~\ref{sec:obs_acrit}.

In terms of $a_{crit,max}$/$a_{crit,init}$, the differences between the two models are modest with the \citet{Repetto2014} magnetic braking model yielding slightly larger values of $a_{crit,max}$/$a_{crit,init}$.  In general, the \citet{Repetto2014} model tends to remove less rotational angular momentum than the model of \citet{Reiners2012}.  The differences between the two models in terms of $a_{crit,max}$/$a_{crit,init}$ decrease with increasing initial stellar $P_{rot}$ as for longer $P_{rot}$ there is less stellar angular momentum for magnetic braking to remove.  The difference between the two models becomes more drastic in the late evolution of the binaries once both stars are tidally locked.  In that case, the slow angular momentum depletion of the \citet{Repetto2014} model causes a slight decay in the orbital period of about a quarter of a day while the \citet{Reiners2012} model produces a decay of 1-1.5 days.  Overall, our results are moderately sensitive to the choice of magnetic braking model and we choose to use the \citet{Reiners2012} model as our fiducial model as it tends to produce a more conservative estimate of $a_{crit,max}$/$a_{crit,init}$.  For both magnetic braking models when the stars have initial $P_{rot} \sim 1$ day, the binary reaches $a_{crit,max}$/$a_{crit,init} \gsim 1.1$, large enough to destabilize CBPs.  

\subsection{Varying Rotation Periods} \label{sec:var_rot}

Previously, we posited that an initial $P_{rot} \lsim 1$ day, a value consistent with observations of young stars (see $\S$~\ref{sec:stellar_rotations}), should be sufficient for the STEEP process to operate.  Here, we demonstrate that initial $P_{rot} \lsim 1$ days do indeed lead to appreciable growth in $a_{crit}$ through an expansive suite of simulations.

\begin{figure*}[t]
	\includegraphics[width=\textwidth]{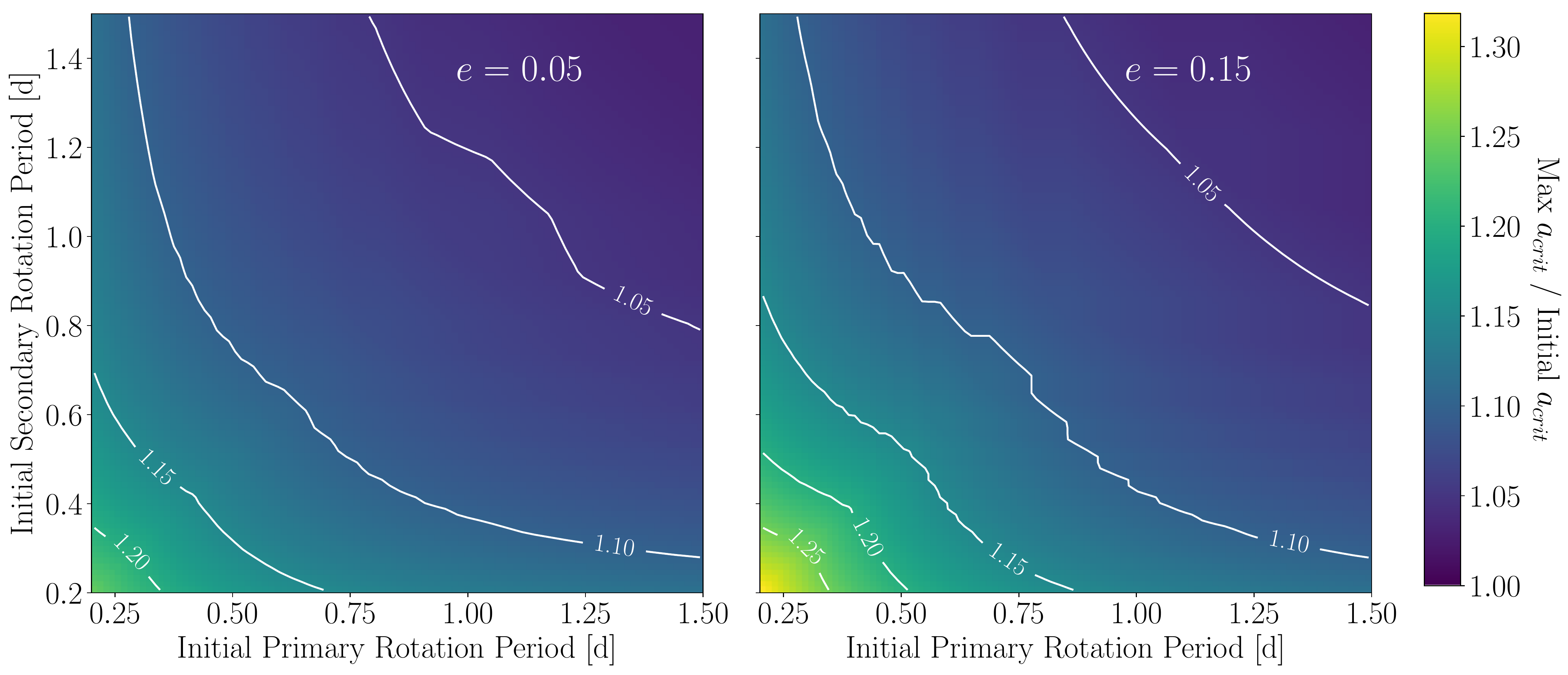}
    \caption{$a_{crit,max}$/$a_{crit,init}$ as a function of the initial stellar $P_{rot}$ for a $1$ M$_{\odot}$$-1$ M$_{\odot}$ binary for $e = 0.05$ ({\it Left}) and $e = 0.15$ ({\it Right}).  The white lines are contours in $a_{crit,max}$/$a_{crit,init}$.  When the stars start with $P_{rot} \lsim 1$ days, $a_{crit,max}$/$a_{crit,init}$ readily grows to $\gsim 1.1$, large enough to destabilize some CBPs near the dynamical stability boundary.}
    \label{fig:GG_contour}
\end{figure*}

In Fig.~\ref{fig:GG_contour}, we present the results of 20,000 simulations in which we varied the initial $P_{rot}$ for both stars over a grid of 0.2-1.5 days for a $1 M_{\odot} - 1 M_{\odot}$ binary for $e=0.05$ and $e=0.15$.  For the other initial conditions, we adopt the default values given in Table~\ref{tab:params}.  For both eccentricities, binaries in which both stars begin with $P_{rot} \lsim 1$ days achieve $a_{crit,max}$/$a_{crit,init} \gsim 1.1$.  When the stars have initial $P_{rot} \lsim 0.5$ days $a_{crit,max}$/$a_{crit,init}$ can reach $1.3-1.4$, large enough to potentially destabilize many {\it Kepler} CBPs as they typically reside at $a_{planet}/a_{crit} \lsim 1.4$.  The initial $P_{rot}$ that lead to appreciable $a_{crit}$ growth are entirely consistent with observations of young sun-like stars, demonstrating that the STEEP process can effectively destabilize some CBPs near the dynamical stability boundary.

\subsection{Monte Carlo Simulations} \label{sec:monte_carlo}

In the previous sections, we ran simulations varying one or two parameters at a time to explore the sensitivity of the STEEP process to the initial conditions.  We found that simulations with parameters roughly consistent with observations, such as initial stellar $P_{rot}$ and parameters broadly consistent with theoretical expectations result in binaries whose coupled stellar and tidal evolution can effectively destabilize CBPs.  However, the coupled nature of our model necessitates a broader study to test the STEEP process's robustness to combinations of the parameters.

\begin{figure}[h]
	\includegraphics[width=\columnwidth]{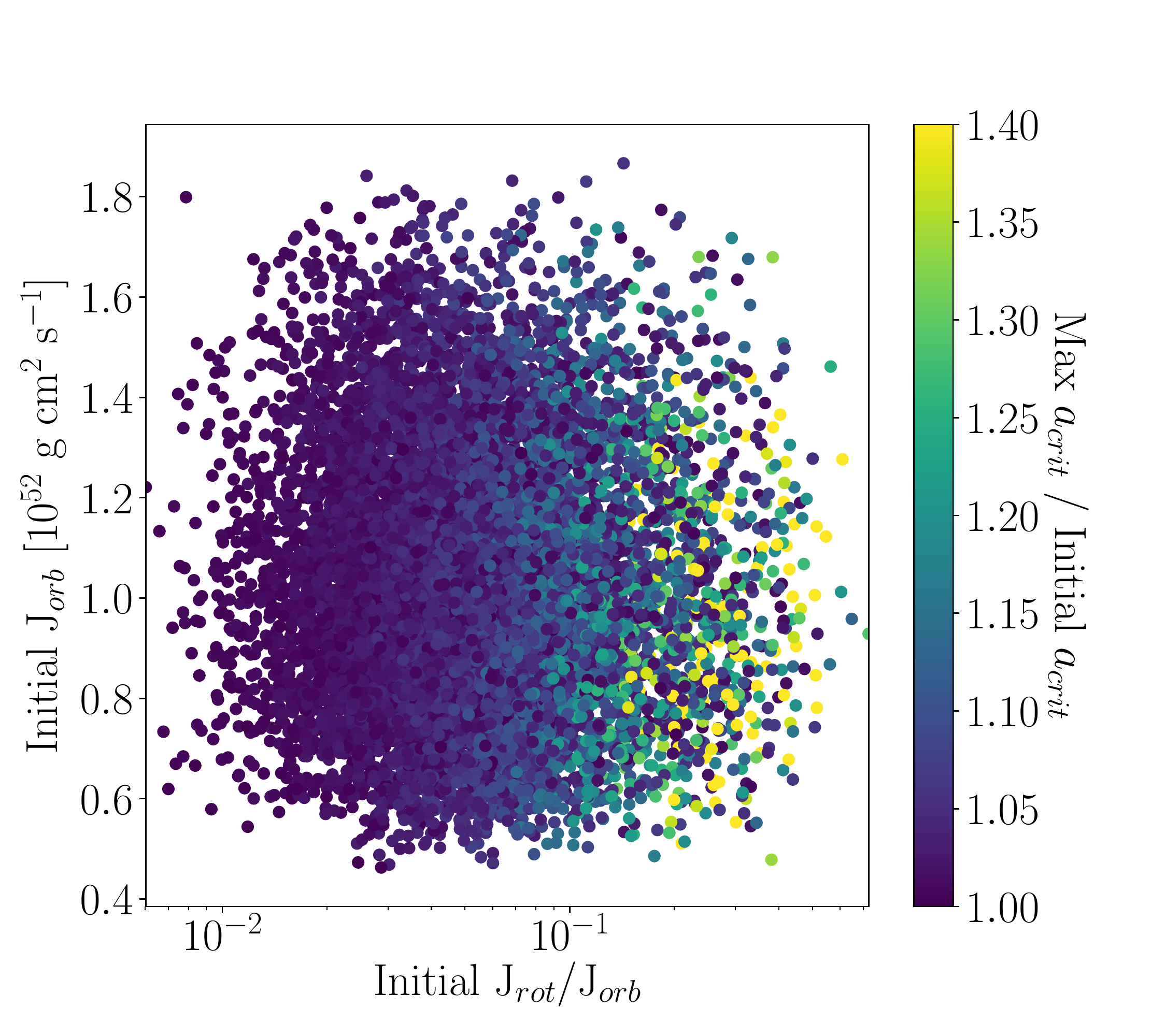}
    \caption{Scatter plot of the initial orbital angular momentum $J_{orb}$, versus the ratio of total initial stellar rotational angular momentum to the initial orbital angular momentum, $J_{rot,tot}$/$J_{orb}$.  Each point is colored by the $a_{crit,max}$/$a_{crit,init}$ achieved in that simulation.  Simulations with an initial angular momentum ratio of ${\gsim}0.1$ tend to produce a large enough $a_{crit,max}$/$a_{crit,init}$ to destabilize a CBP.}
    \label{fig:mc_uniform}
\end{figure}

We therefore perform a Monte Carlo study to identify regions in parameter space where large $a_{crit,max}$/$a_{crit,init}$ occur.  We run 10,000 simulations where each star's mass, $r_g$, initial $P_{rot}$, initial binary orbital period and $e$ are randomly sampled from uniform distributions from the ranges listed in Table~\ref{tab:params}.  Each star's tidal Q was sampled randomly from a log-uniform distribution from the listed range.  For each simulation, we compute $a_{crit,max}$/$a_{crit,init}$ from the full evolution of the system over 2 Gyr.  The results of the simulations are displayed in Fig.~\ref{fig:mc_uniform} as a two-dimensional projection in terms of the initial orbital angular momentum, $J_{orb}$, and ratio of total initial stellar rotational angular momentum to the initial orbital angular momentum, $J_{rot,tot}$/$J_{orb}$.

A clear gradient emerges in $J_{rot,tot}$/$J_{orb}$ space.  As $J_{rot,tot}$/$J_{orb}$ increases, on average so does $a_{crit,max}$/$a_{crit,init}$.  This result is expected since when the initial rotational angular momentum content is comparable to the orbital angular momentum, tidal transfer to the orbit will on average increase $a_{crit}$ more.  We find that simulations with initial $J_{rot,tot}$/$J_{orb} \gsim 0.1$ yield $a_{crit,max}$/$a_{crit,init} \gsim 1.1$, large enough to destabilize CBPs near the dynamical stability boundary.  Simulations with initial $J_{rot,tot}$/$J_{orb} \approx 1$ can result in simulations with large $a_{crit,max}$/$a_{crit,init} \approx 1.4$, some upwards of $2-3$ for initial stellar rotation rates near the break-up velocity, a value that is almost certain to destabilize and eject any nascent circumbinary planetary system.

There is considerable scatter in $a_{crit,max}$/$a_{crit,init}$, however, as simulations with $J_{rot,tot}$/$J_{orb} \approx 1$ can result in little evolution in $a_{crit}$.  We find that these cases correspond to simulations where one or both stars have tidal Qs ${\sim}10^7$.  As discussed in $\S$~\ref{sec:var_Q}, large tidal Qs result in much slower tidal evolution.  When the tidal evolution proceeds more slowly, so too does the angular momentum transfer into the orbit (see Fig.~\ref{fig:var_Q}).  In this case, the slow tidal evolution allows for magnetic braking to efficiently siphon large amounts of angular momentum from the system, resulting in negligible $a_{crit}$ evolution.

For tidal Qs low enough to prevent significant angular momentum loss due to magnetic braking, i.e. $Q < 10^7$, young binary systems with short initial $P_{rot} \lsim 1$ day can attain $J_{rot,tot}$/$J_{orb} \gsim 0.1$, resulting in large enough growth in $a_{crit}$ to destabilize CBPs found near the dynamical stability boundary.  

\subsection{3:2 Spin-Orbit Resonance} \label{sec:32}

\begin{figure}[h]
	\includegraphics[scale=0.5]{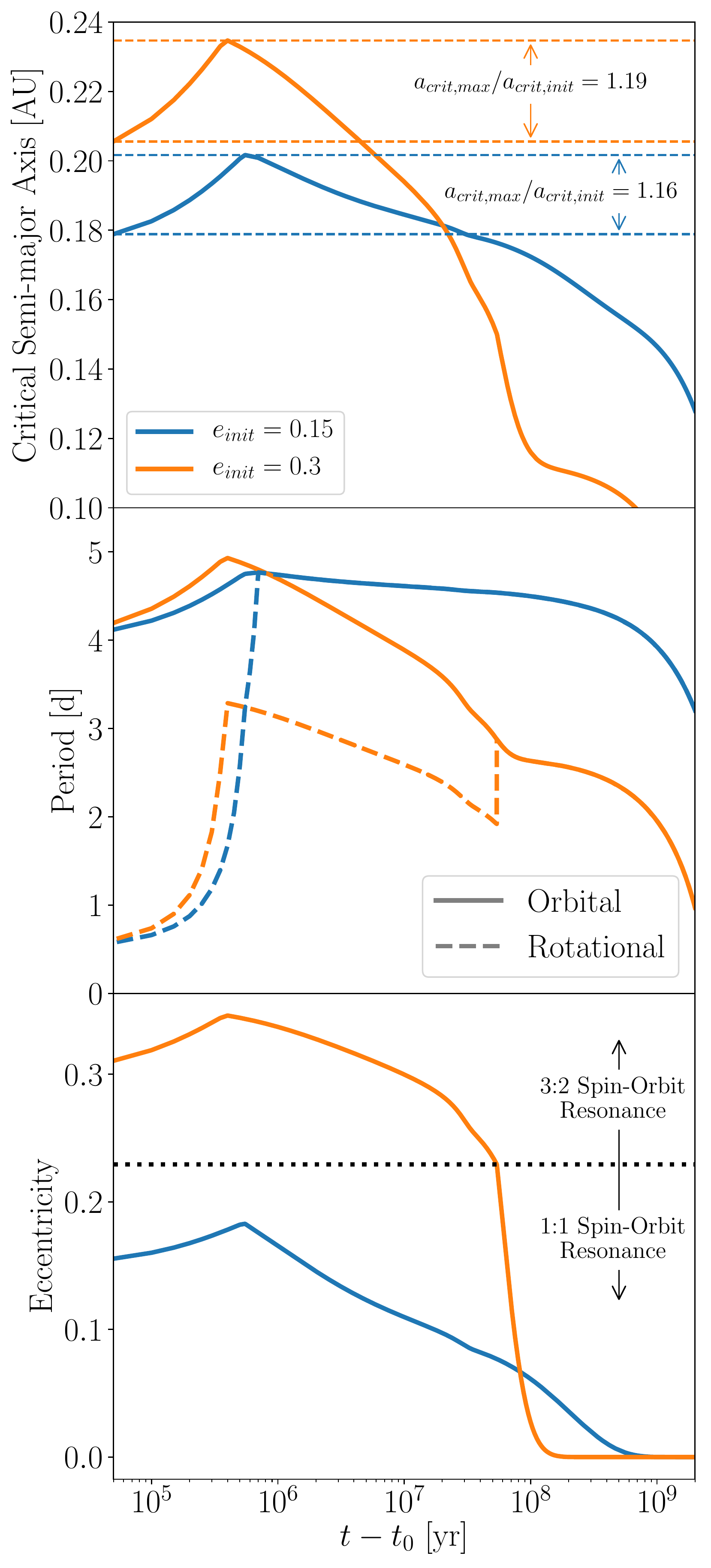}
   \caption{Full evolution of binary systems using our default parameters from Table~\ref{tab:params}, except with initial eccentricities of $e = 0.15$ (blue) and $e = 0.3$ (orange).  {\it Top:} $a_{crit}$ vs. time.  The systems attain $a_{crit,max}/a_{crit,init} = 1.16, 1.19$ for binaries with initial eccentricities of $e = 0.15$ and $e = 0.3$, respectively. {\it Middle:} Orbital and stellar rotation period vs. time.  Both systems tidally lock after ${\sim}1$ Myr into 1:1 (blue) and 3:2 (orange) spin-orbit resonances.  {\it Bottom:} $e$ vs. time.  The dotted line demarcates binaries that tidally locked into a 3:2 (above the line) and 1:1 (below the line) spin-orbit resonance.}
    \label{fig:ecc_comp}
\end{figure}

To probe the impact of large $e$ and capture into higher order spin-orbit resonances, we run two simulations using the default parameters in Table~\ref{tab:params} but with $e = 0.15$ and $e = 0.3$.  The results of the simulations are given in Fig.~\ref{fig:ecc_comp}.  The initially more eccentric binary tidally locks into a 3:2 spin-orbit resonance after about 1 Myr since $e > \sqrt{1/19}$ (see $\S$~\ref{sec:tidal_evolution}), while the less eccentric binary tidally locks into and remains in a synchronous 1:1 spin-orbit state.  Once $e$ decays to $e < \sqrt{1/19}$ after about 100 Myr for the more eccentric binary, the system becomes trapped in the synchronous state, the only other allowed spin-orbit resonance for tidally locked systems under the CPL model.

The binary that tidally locks into a 3:2 spin-orbit resonance reaches a larger $a_{crit,max}/a_{crit,init}$ than the other synchronously rotating binary system.  This result is surprising given that in a 3:2 spin-orbit resonance, $P_{rot} = 2/3 P_{orb}$, so relative to the synchronous case, less stellar rotational angular momentum is transported into the orbit, reducing the increase in orbital period and hence $a_{crit}$.  However, larger $e$ correspond to lower orbital angular momentum which scales as $\sqrt{1-e^2}$.  The more eccentric binary system has less orbital angular momentum than the other system so transfer of angular momentum from the stellar rotation to the orbit yields proportionally larger increases in $a_{crit}$, explaining the larger $a_{crit,max}/a_{crit,init}$.  Therefore we anticipate that the STEEP process is effective for eccentric binary systems that tidally lock into higher order spin-orbit resonances.

\subsection{Long-term $a_{crit}$ Evolution} \label{sec:obs_acrit}

As shown in the above simulations, $a_{crit}$ changes over time such that a binary's observed $a_{crit}$ can differ significantly from past values.  This effect is especially relevant for observed {\it Kepler} circumbinary systems as the orbit, and hence $a_{crit}$, observed today is much different than in the past and the magnitude of the difference depends on the age of the system.  Most notably, the post-tidal locking $a_{crit}$ decay implies that the dynamical instability region around the binary was likely larger in the past.  Any CBPs that survived initial $a_{crit}$ increases would necessarily appear to orbit on much larger $a_{CBP}$ relative to the $a_{crit}$ about the central binary and would therefore be harder to detect.

\subsubsection{Specific Cases}

\begin{figure*}[t]
	\includegraphics[width=\textwidth]{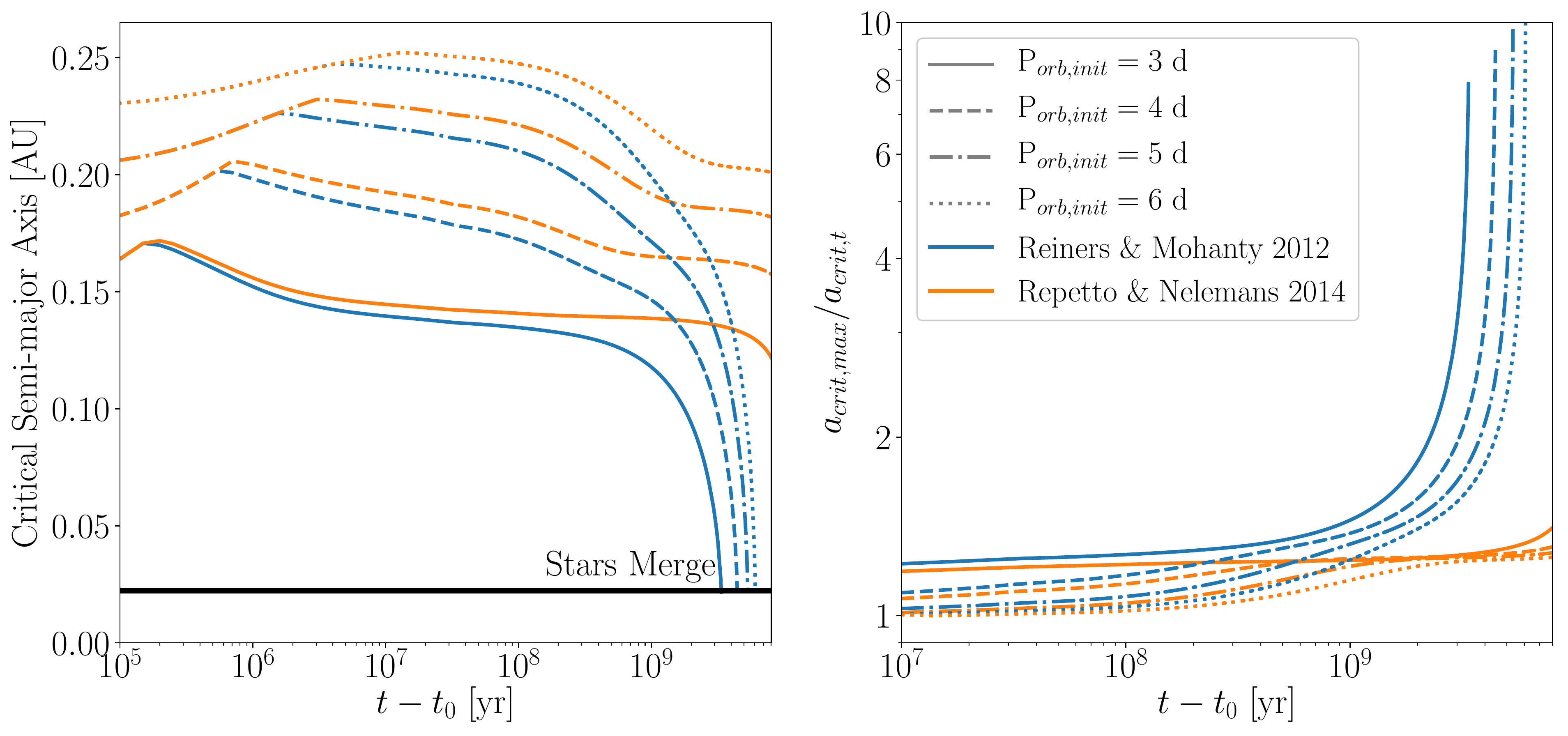}
   \caption{{\it Left:} $a_{crit}$ as a function of time.  The black horizontal line indicates when stellar radii overlap and the stars merge, halting the simulation. {\it Right:} Maximum critical semi-major axis divided by the critical semi-major axis observed at time $t$, $a_{crit,max}/a_{crit,t}$.  Depending on the magnetic braking physics, the observed $a_{crit}$ can differ from its maximum by an order of magnitude.}
    \label{fig:obs_acrit}
\end{figure*}

To illustrate this effect and the observational consequences, we simulate 4 binary systems for 8 Gyr of evolution using the default parameters in Table~\ref{tab:params} for binary orbital periods of 3, 4, 5, and 6 days using the magnetic braking formalisms of \citet{Reiners2012} and \citet{Repetto2014}.  The results of the simulations are shown in Fig.~\ref{fig:obs_acrit}.

In left panel of Fig.~\ref{fig:obs_acrit}, we see the full evolution of these systems and there is a stark difference between the two cases: the \citet{Reiners2012} magnetic braking law causes significant orbital decay relative to the modest decay induced by the \citet{Repetto2014} magnetic braking relation.  The \citet{Reiners2012} magnetic braking law removes a significant amount of angular momentum from the orbits driving orbital period decays of about 1 day per Gyr such that the binaries actually merge after a few Gyr as the stellar radii overlap.  The orbital period decay is more modest for the \citet{Repetto2014} magnetic braking law.  The different outcomes predicted by these models could be used as an observational test to constrain which magnetic braking model is more suitable, however, we leave that analysis for future work.

In the right panel of Fig.~\ref{fig:obs_acrit}, we plot $a_{crit,max}/a_{crit}$ over time.  For the simulations using the \citet{Reiners2012} magnetic braking law, $a_{crit}$ can vary by a factor of $2-10$ from its maximum value depending on the age of the system, while the simulations using the \citet{Repetto2014} law vary by factors of order unity.  This implies that the region of dynamical instability can appreciably shrink for tidally locked short orbital period binaries over time.  Therefore when one observes these systems, the fact that the dynamical instability region was likely much larger in the past precludes any CBPs from orbiting near the observed $a_{crit}$.  Any CBPs that survived the $a_{crit}$ increase would then be orbiting at much larger $a_{\text{CBP}}/a_{crit}$ ratios than what is observed for {\it Kepler} CBPs.  If a CBP is discovered around a tidally locked binary in the future, its current state must be considered in the context of the host binary's past coupled stellar-tidal evolution.

\subsubsection{Monte Carlo Simulations}

\begin{figure}[h]
	\includegraphics[width=\columnwidth]{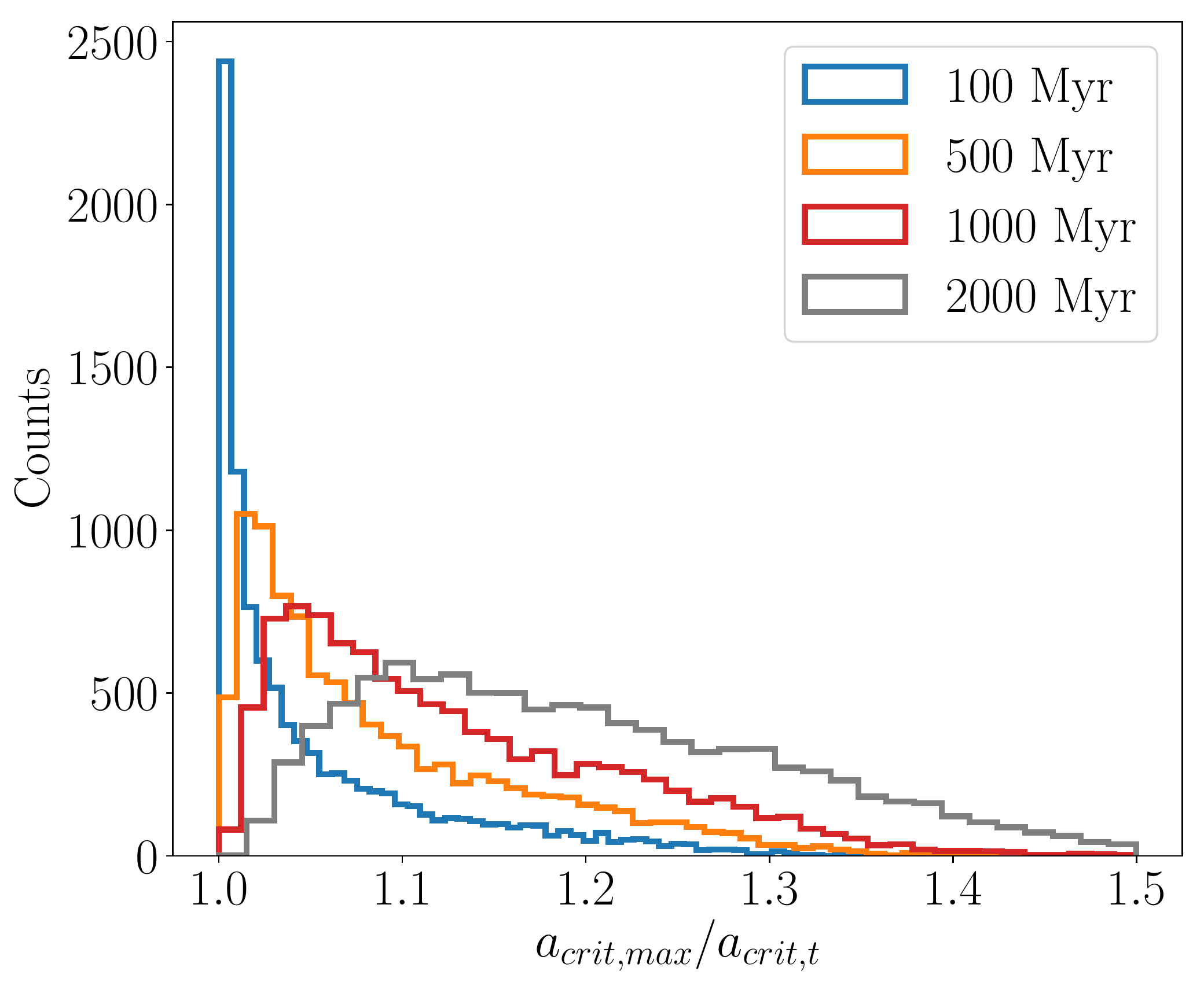}
    \caption{Histograms of the maximum $a_{crit}$ divided by the critical semi-major axis observed at time $t$, $a_{crit,max}/a_{crit,t}$ for the binary systems simulated in $\S$~\ref{sec:monte_carlo} observed at 100, 500, 1000, and 2000 Myr (blue, orange, red, and grey histograms, respectively).  In general for older systems, $a_{crit,max}/a_{crit,t}$ increases as $a_{crit}$ recedes due to angular momentum loss via magnetic braking in tidally locked systems.  The distribution of $a_{crit,max}/a_{crit,t}$ smears out and develops an extended tail for older systems due to the different tidal and rotational properties of the binary systems.}
    \label{fig:acrit_hist}
\end{figure}

Here we examine how distributions of $a_{crit,max}/a_{crit,t}$ from our simulations from $\S$~\ref{sec:monte_carlo} evolve as a function of time to examine the impact of different initial conditions.  We plot the distribution of $a_{crit,max}/a_{crit,t}$ at four different times in Fig.~\ref{fig:acrit_hist}.  As the systems age, the distributions tend to shift towards larger $a_{crit,max}/a_{crit,t}$ values, indicating that for most initial states $a$, and hence $a_{crit}$, decays relative to its peak due to magnetic braking.  At later times, the $a_{crit,max}/a_{crit,t}$ distributions tend to smear out and a develop heavy tail indicating that even though $a_{crit,max}/a_{crit,t}$ tends to grow, the properties on the individual system, e.g. the stellar mass or tidal Qs, play an important role in determining how much $a_{crit}$ evolves.  For example, lower mass stars contract for longer periods of time, which for tidally locked stars, injects additional angular momentum in the orbit, slowing $a_{crit}$ decay due to magnetic braking.

\subsection{Relaxed Assumptions} \label{sec:optimistic}

In previously-discussed simulations, we have taken a rather conservative approach in selecting our initial conditions.  Here we relax those assumptions and run five simulations with progressively more extreme-yet-plausible initial conditions to see how $a_{crit}$ evolves. Case A has the same initial conditions as our fiducial simulation (see $\S$~\ref{sec:fiducial_simulation}) but with an initial orbital period of 5 days and an initial $e$ of 0.1.  Case B is the same as Case A but with both stellar tidal Qs set to $10^5$.  Case C is the same as Case B but with both stellar radii of gyration set to $r_g = 0.45$.  In Case D, the initial conditions are the same as Case C but with the initial stellar $P_{rot}$ set to $0.25$ days.  Finally in Case E, we set the initial $e=0.2$ and adopt the initial conditions of Case D for all other parameters.

\begin{figure*}[t]
	\includegraphics[width=\textwidth]{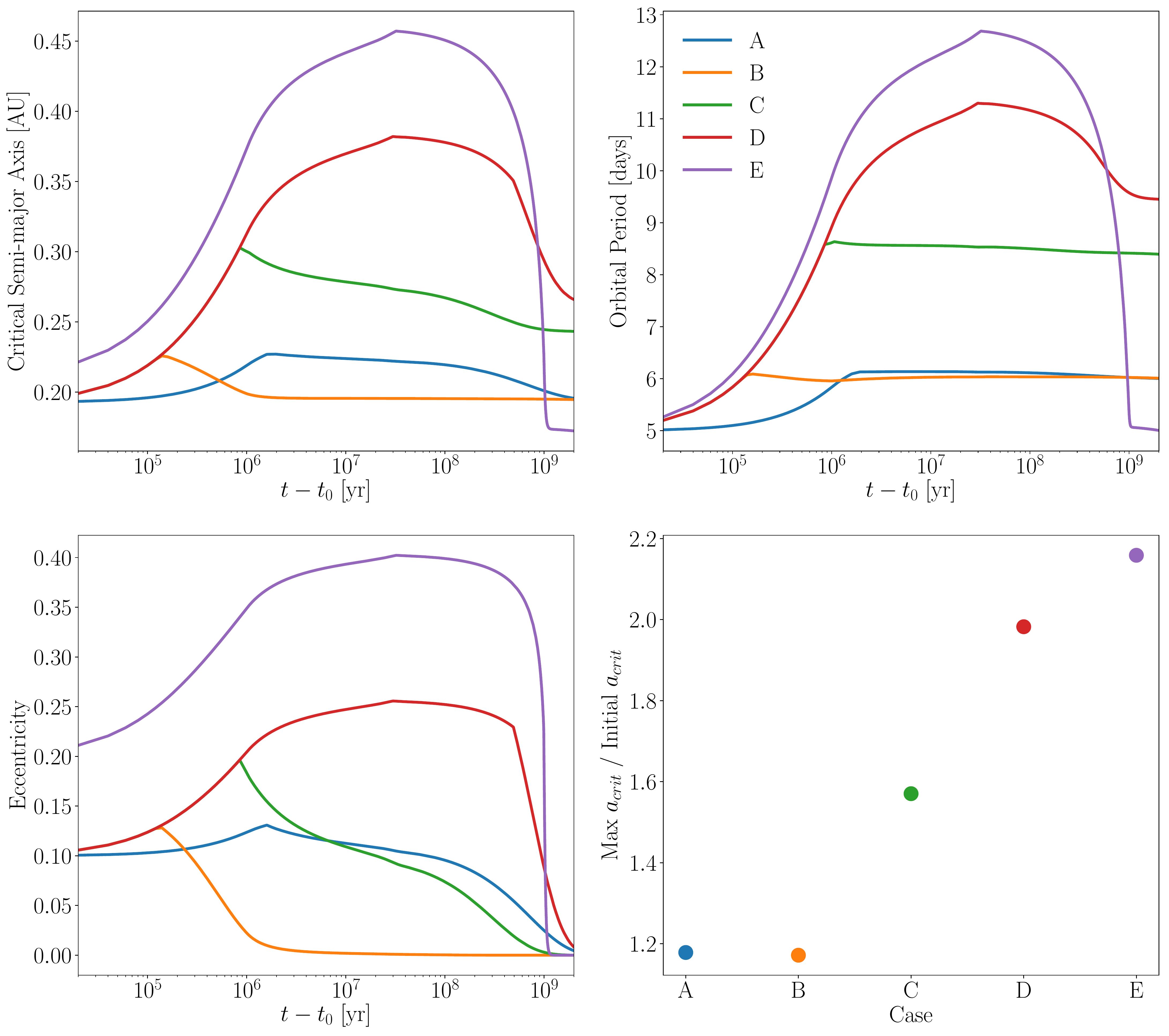}
   \caption{Binary orbital evolution for several plausible sets of initial conditions. {\it Top Left:} $a_{crit}$ vs. time.  {\it Top Right:} Binary orbital period vs. time. {\it Bottom Left:} $e$ vs. time. {\it Bottom Right:} $a_{crit,max}$/$a_{crit,init}$ for the five cases.  The blue, orange, green, red, and purple curves correspond to cases A, B, C, D, and E, respectively.  In agreement with previously discussed simulations, we find that initial fast rotators with smaller tidal Qs achieve large $a_{crit,max}$/$a_{crit,init}$ up to of order $2$ for plausible initial conditions.}
    \label{fig:opt}
\end{figure*}

For each individual simulations, see Fig.~\ref{fig:opt}, the results proceed as expected: simulations with larger $r_g$ and lower initial $P_{rot}$, e.g. Case C and Case D, respectively, result in larger $a_{crit,max}$/$a_{crit,init}$ with the ratios approaching 2.  In all simulations, the $a_{crit}$ growth peaks early on in the system at $\lsim 100$ Myr.  For binaries with $e > 0.3$, the model tends to break down and yield unrealistic results and merits further exploration with the CTL tidal model, which may be applicable at large $e$.  Nevertheless, these simulations demonstrate that $a_{crit}$ can grow up to twice the initial value depending on the initial conditions.

\section{Results: N-body Simulations} \label{sec:nbody_results}

Here we present the results of N-body simulations of circumbinary planetary systems using the N-body code \texttt{REBOUND} \citep{Rein2012,Rein2015}.  The initial conditions and set-up are described in $\S$~\ref{sec:nbody_initial_conditions}.  In $\S$~\ref{sec:nbody_dynamics}, we examine how orbital instabilities in circumbinary planetary systems stemming from the inner-most planet residing within $a_{crit}$ affect the system architecture and lead to planetary ejections while in $\S$~\ref{sec:nbody_observation}, we examine the observational consequences of CBP ejections.

\subsection{Dynamical Stability} \label{sec:nbody_dynamics}

\subsubsection{Single Planet Circumbinary System} 

In Table~\ref{tab:nbody_single}, we display the fraction of simulations of a single planet circumbinary system that result in a stable or unstable planet for Neptune-, Saturn-, and Jupiter-mass planets.  In our simulations, planet b, which initially resided within $a_{crit}$, is ejected $70\%$ - $75\%$ of the time confirming that the majority of CBPs that drift interior to $a_{crit}$ are ejected from the system \citep[c.f.][]{Holman1999}.  From the definition of $a_{crit}$ from Eq.~(\ref{eqn:crit_semi}), however, one expects \textit{every} CBP that drifts interior to $a_{crit}$ to go unstable and get ejected whereas only $70\%$ - $75\%$ of such planets are ejected in our simulations.  The difference can be rectified by examining precisely how  \citet{Holman1999} computed $a_{crit}$.  For each given $a_{CBP}$, \citet{Holman1999} initialized 8 test particles on circular orbits that are equally spaced in mean anomaly.  After the integration, \citet{Holman1999} deemed the minimum semi-major axis in which all 8 test particles survive $a_{crit}$.  Therefore a test particle, and hence a planet, within $a_{crit}$ is likely, but not guaranteed, to go dynamically unstable and be ejected from the system, explaining our results.

We find that on average, the deeper within the region of dynamical instability a CBP is and the more eccentric its orbit, the more likely it is to be ejected from the system.  There appears to be a weak dependence on ejection probability with planet mass with Neptune-mass CBPs getting ejected ${\sim}75\%$ of the time while more massive planets were slightly less likely to be ejected.  We re-ran these simulations for $10^6$ binary orbital periods, an order of magnitude longer, and found that our results did not significantly change.

\begin{deluxetable}{lcc}
\tabletypesize{\small}
\tablecaption{Single Planet System Outcome Fraction \label{tab:nbody_single}}
\tablewidth{0pt}
\tablehead{
\colhead{Case} & \colhead{Stable} & \colhead{Unstable}
}
\startdata
Neptune Mass & 0.253 & 0.747 \\
Saturn Mass & 0.282 & 0.718 \\
Jupiter Mass & 0.3 & 0.7
\enddata \vspace*{0.1in}
\end{deluxetable}

\subsubsection{Multiplanet Circumbinary System}

In Table~\ref{tab:nbody_multi}, we display the fraction of simulations of a two-planet circumbinary system that result in a stable or unstable planet as a function of initial planet orbital inclination distribution and planet mass. A clear result of our multi-planet circumbinary system dynamical stability simulations is that at least one planet is ejected from the system with this result occurring in $87\%$ - $95\%$ of simulations, a value broadly consistent with the results of previous studies of planet-planet scattering in circumbinary systems \citep[e.g.][]{Sutherland2016,Smullen2016,Gong2017,Gong2017b}. 

In simulations in which a planet is ejected, the typical result is planet c remaining stable while planet b is ejected from the system.  This outcome occurs in $78\%$ - $88\%$ of simulations.  Planet b remains stable $7\%$ - $18\%$ of the time where most instances in which b remains stable correspond to simulations in which both planets remain stable.  Simulations in which both b and c remain stable typically correspond to initial conditions in which b and c are more widely separated with inter-planet separations near $10 R_{Hill,mutual}$ and when b resides near to $a_{crit}$.  The few simulations in which b remains stable while c is ejected correspond to scattering events where b scatters exterior to c while c scatters into the region of dynamical instability and is ejected from the system soon thereafter.  Planet b remains stable at the expense of c's ejection in $2\%$ - $6\%$ of simulations.  The spread in simulation outcomes stems from the planet(s)' and binaries' random initial orbital parameters (see Table~\ref{tab:nbody_params}).

The most infrequent result that occurs in $1\%$ - $5\%$ of simulations is when both b and c are ejected from the system, a value roughly consistent with \citet{Gong2017b} who find ${\sim 10\%}$ systems are destabilized due to planet-planet scattering near $a_{crit}$.  Simulations in which both b and c are ejected tend to occur when b is initialized deeper within the region of dynamical instability and when b and c are more closely separated with separations near $5 R_{Hill,mutual}$ in good agreement with the simulations of \citet{Kratter2014}.  Given that planet c survives in $89\%$ - $97\%$ of simulations, we anticipate that in systems with higher multiplicity, the farther out planets would likely remain stable.

Our results show clear dependences on both CBP orbital inclination relative to the binary's and planet's mass.  Systems with more massive planets tend to remain stable. For a given planet mass, however, simulations with planets initialized using the ``high inclination" distribution more frequently result in planet b going unstable as the larger mutual inclinations tend to result in more violent scattering events.  This behavior is reflected in a depletion in the fraction of simulations in which b remains stable and an appreciable enhancement in the fraction of simulations in which planet c remains stable after b is ejected.

\begin{deluxetable*}{lcccccc}[th]
\tablewidth{\linewidth}
\tablecaption{Circumbinary planetary system N-body simulation outcome fractions.  \label{tab:nbody_multi}}
\tablehead{
\colhead{} &  \multicolumn{3}{c}{Low Inclination} & \multicolumn{3}{c}{High Inclination} \\
\cline{2-4} \cline{5-7} \\
\colhead{Case} & \colhead{Neptune Mass} & \colhead{Saturn Mass} & \colhead{Jupiter Mass} & \colhead{Neptune Mass} & \colhead{Saturn Mass} & \colhead{Jupiter Mass}
}
\startdata
b, c stable  & $0.064$ & $0.088$ & $0.132$ & $0.053$  & $0.09$ & $0.106$ \\
b stable, c unstable & $0.063$ & $0.047$ & $0.052$  & $0.027$  & $0.017$  & $0.023$ \\
c stable, b unstable & $0.825$ & $0.843$  & $0.784$  & $0.881$  & $0.879$  & $0.849$ \\
b, c unstable & $0.048$ & $0.022$  & $0.032$  & $0.039$  & $0.014$  & $0.022$
\enddata
\tablecomments{In ``low inclination" simulations, both planet b and c's inclination with respect to the binary orbital plane is sampled from $U(0^{\circ},1^{\circ})$ while in ``high inclination" simulations, both planet b and c's inclination is sampled from $U(0^{\circ},3^{\circ})$.  The fractions are normalized such that each column sums to $1$.}
\end{deluxetable*}

Our results indicate that systems in which the inner-most CBP falls within the region of dynamical instability, \textit{e.g.} due to $a_{crit}$ expansion resulting from coupled stellar-tidal evolution, likely lose at least one planet.  Therefore for short-period binary systems where we expect coupled stellar-tidal evolution to increase $a_{crit}$ and envelope CBPs that preferentially lie near the limit, at least one CBP is likely to be ejected from the system, potentially accounting for the lack of observed CBPs in such systems.

We note that our results are conservative given our assumptions that the inner-most CBP orbits just interior to $a_{crit}$ to simulate the time right after the planet falls within $a_{crit}$ due to the STEEP process. In practice, a CBP that falls within $a_{crit}$ and is not promptly ejected will fall deeper into the dynamical instability region as coupled stellar-tidal evolution expands $a_{crit}$.  As discussed above, the deeper a CBP is within the dynamical instability region, the more likely it is to be ejected.  Therefore if we relaxed our conservative assumptions and initialized CBPs more interior to $a_{crit}$, we would expect more ejections.  Additionally it is possible that CBPs form with large mutual inclinations although the observed CBP population has low mutual inclinations with respect to the host binaries \citep{Li2016}.  If we allowed for more inclined orbits, based off of our previous results we would anticipate more ejections based on the results of \citet{Chatterjee2008} and more CBPs scattered away from a transiting configuration.  

\subsection{Mock Transit Observations} \label{sec:nbody_observation}

\begin{figure*}[t]
	\includegraphics[width=\textwidth]{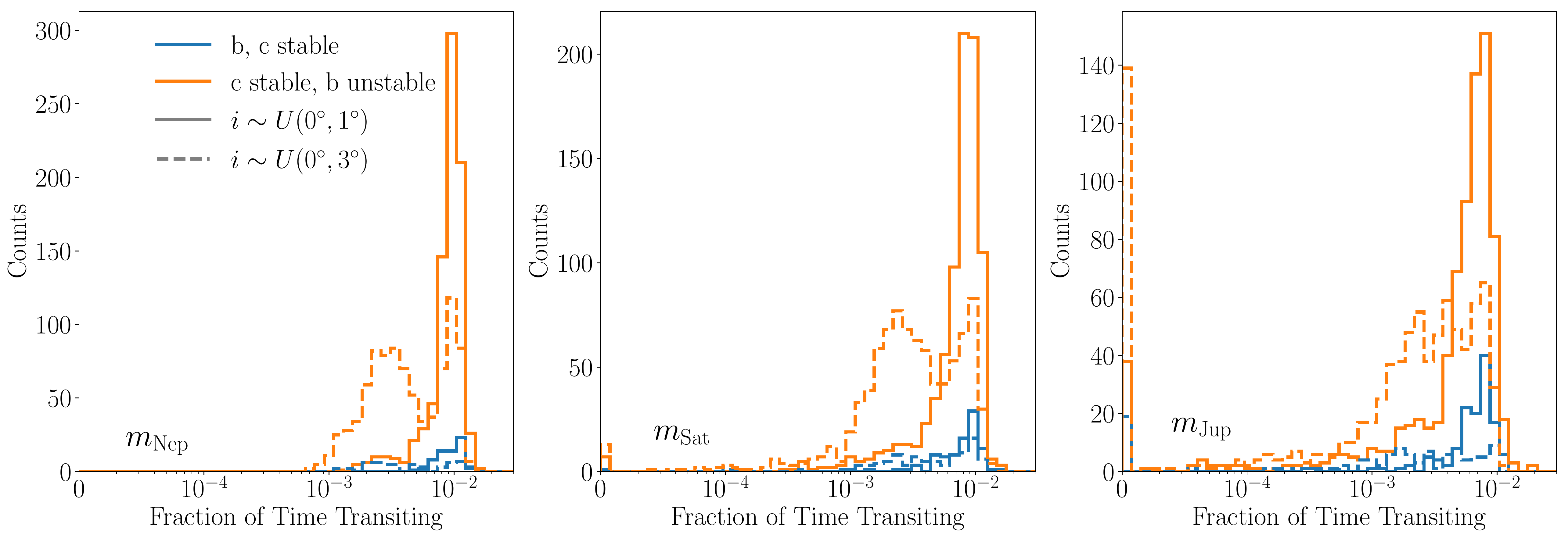}
    \caption{Histogram of planet c's fraction of time spent transiting one of the binary stars for Neptune- (\textit{Left:}), Saturn- (\textit{Middle:}), and Jupiter-mass (\textit{Right:}) exoplanets.  The orange histograms correspond to when c remains stable while b goes unstable and the blue curve corresponds to when both b and c remain stable.  The solid and dashed histograms correspond to cases where both planets' initial inclinations are uniformly sampled from $[0^{\circ},1^{\circ}]$ and $[0^{\circ},3^{\circ}]$, respectively.  Generally, the larger the mutual inclination, the less time c spends transiting.}
    \label{fig:nbody_transit}
\end{figure*}

Next we present the results of our mock transit observation simulations.  In the single CBP system case, the result is trivial: if the planet is ejected, it does not transit while if it remains in the system, it most likely transits as each planet was initialized in a transiting configuration and interactions with the host binary tend to not excite large inclinations given the small initial mutual inclination.  For the two-planet circumbinary system case, we focus on the detectability of planet c as we have shown that the dominant outcome of our dynamical stability N-body simulations is that planet b is ejected while planet c remains in the system on a perturbed-yet-stable orbit.  We summarize the results of our mock transit observations in Fig.~\ref{fig:nbody_transit} which depicts a histogram of the fraction of time transiting (FTT) for planet c for various initial masses and inclinations for the case in which both planets b and c remain stable (blue lines) and the case in which planet b gets ejected while planet c remains (orange lines).  For reference, a Jupiter-sized exoplanet with $i=90^{\circ}$ on a 30 day circular orbit, a typical orbital period for planets in our simulations, orbiting a Sun-like star spends about $1\%$ of the time transiting for a FTT of $10^{-2}$.  Almost all of the CBPs in our simulations are initialized in a transiting configuration and typically have an initial FTT of order $10^{-2}$.  

We find that in most cases in which planet b is ejected while planet c remains stable, planet b's ejection does little to impact the ``transitability" of the remaining planet.  In the case of Neptune-mass planets in Fig.~\ref{fig:nbody_transit}, the histograms for b unstable, c stable and b, c stable have similar shapes indicating that planet b's ejection does little to change planet c's FTT.  For Saturn-mass planets, specifically those with an initially larger inclination, there is a small extended tail in the FTT distribution for the case when planet b is ejected while planet c remains stable indicating that more inclined, massive planets are scattered to slightly larger inclinations and eccentricities, reducing their FTT.  

From Fig.~\ref{fig:nbody_transit}, clear trends in both mass and initial inclination can be seen.  After an instability event occurs in which planet b is ejected, the more massive the CBPs, the less time they spend transiting.  For Neptune-mass CBPs after planet b is ejected, planet c's orbit does not appreciably change and retains a large FTT, see the left panel of Fig.~\ref{fig:nbody_transit}.  As the planet mass increases, planet b's chaotic evolution significantly perturbs planet c's orbit, scattering it to larger inclinations, reducing FTT.  This effect is most notable in the extended tail and large peak at FFT near 0 for the orange lines in the right panel of Fig.~\ref{fig:nbody_transit} for Jupiter-mass planets.  This demonstrates that planet b's ejection can readily scatter planet c away from a transiting configuration for massive planets, preventing its detection via the transit method.  This effect is less pronounced for less massive CBPs.

The initial inclination significantly impacts the subsequent CBP transitability.  The more inclined planet c initially is, the lower the FTT as planet c in general has a larger initial impact parameter and can scatter more violently to higher inclinations when planet b is ejected from the system.  This effect is most clearly seen in the right panel of Fig.~\ref{fig:nbody_transit} for Jupiter-mass CBPs as there is a prominent peak near $\text{FTT}= 0$ for the ``high inclination" case (dashed line) that is much larger than the $\text{FTT}=0$ peak for the ``low inclination" case (solid line).  This difference indicates that more highly-inclined Jupiter-mass CBPs are preferentially scattered away from a transiting configuration compared to less-inclined planets.  In addition to the slight increase in inclination for some surviving planets reducing their FTT, we find that $a_c$ also increases slightly after planet b's ejection, reducing its FTT.  This finding is in good agreement with \citet{Gong2017}.

Our results indicate that when the inner-most planet in a multi-planet circumbinary system is ejected after falling within $a_{crit}$, it is likely that there will be little impact on the remaining planet's detectability.  In the case of massive planets and/or large initial inclinations, however, the inner-most planet's ejection can more readily scatter the remaining planet away from and potentially out of a transiting configuration.  Even in this case, however, there is likely little to no change in the surviving planet's FTT.  Of course, more distant planets are in general less likely to transit, so the STEEP process still reduces the total number of transiting CBPs.

One complicating factor is that we make our mock transit observations right after the dynamical stability integrations.  From our coupled stellar-tidal \vplanet simulations, we showed that the maximum $a_{crit}$ value is typically achieved well within 1 Gyr (see $\S$~\ref{sec:results}) indicating that if a CBP will be enveloped by $a_{crit}$ and fall within the region of dynamical instability, it will happen early on in the system's lifetime.  Our mock transit observations implicitly assume that we observe the system soon after $a_{crit}$ has enveloped the inner-most planet in a given system, a time that likely occurs well within the 1st Gyr of the system's lifetime.  In real transit survey's like the \textit{Kepler} mission, the observed stars are not likely to have ages $\leq 1 \text{Gyr}$ like the systems in our mock observations.  As shown in $\S$~\ref{sec:obs_acrit} for tidally locked binary star systems, the older a system is, the more the binary semi-major axis and hence $a_{crit}$ has decayed leaving the surviving CBPs to to orbit at larger $a_{CBP}/a_{crit}$ than they did in the past.  This effect complicates the detection of CBPs via the transit method and requires that any CBPs discovered around short-period binaries must be understood in the context of the host binary's past coupled stellar-tidal evolution.


\section{Application to Kepler-47} \label{sec:kepler47}

We apply the STEEP process to Kepler-47, the shortest period planet-hosting binary system \citep{Orosz2012}.  Kepler-47 is a nearly circular G and M dwarf binary with an orbital period of about $7.45$ days.  Kepler-47 is the only known multi-planet hosting binary with three CBPs, all with nearly co-planar, low-eccentricity orbits \citep{Orosz2012,Welsh2015}.  Kepler-47 appears to have undergone appreciable tidal evolution as the primary star's $P_{rot}$ is about $4\%$ longer than the orbital period, an indication that the binary is near tidal synchronization \citep{Orosz2012}.  Kepler-47 is a suitable candidate for the STEEP process so we seek to examine how its planets could have survived a potential destabilization from the coupled stellar-tidal evolution of the binary.  

The planets in the Kepler-47 system are all of order Neptune mass or less, so we can draw comparisons with our two-planet CBP system N-body simulations of Neptune mass planets in $\S$~\ref{sec:nbody_results} (see also Fig.~\ref{fig:nbody_transit}).  As shown previously, the ejection of the inner-most planet in a low-mass CBP system typically results in little or no change in the orbits and hence transitability of the surviving exterior planets.  Although rarely the inner-most planet's ejection can destabilize the entire system, the dominant result that the rest of the system remains stable and effectively unchanged.  If the Kepler-47 system did previously have a planet interior to Kepler-47b that was ejected after falling into the region of dynamical instability, it could have been ejected from the system without any noticeable observational impact on the surviving planetary system.  With three planets on nearly co-planar and low-eccentricity orbits, the Kepler-47 planetary system seems relatively dynamically cold, supporting this picture and making its current state compatible with the STEEP process.

\begin{figure}[t]
	\includegraphics[width=\columnwidth]{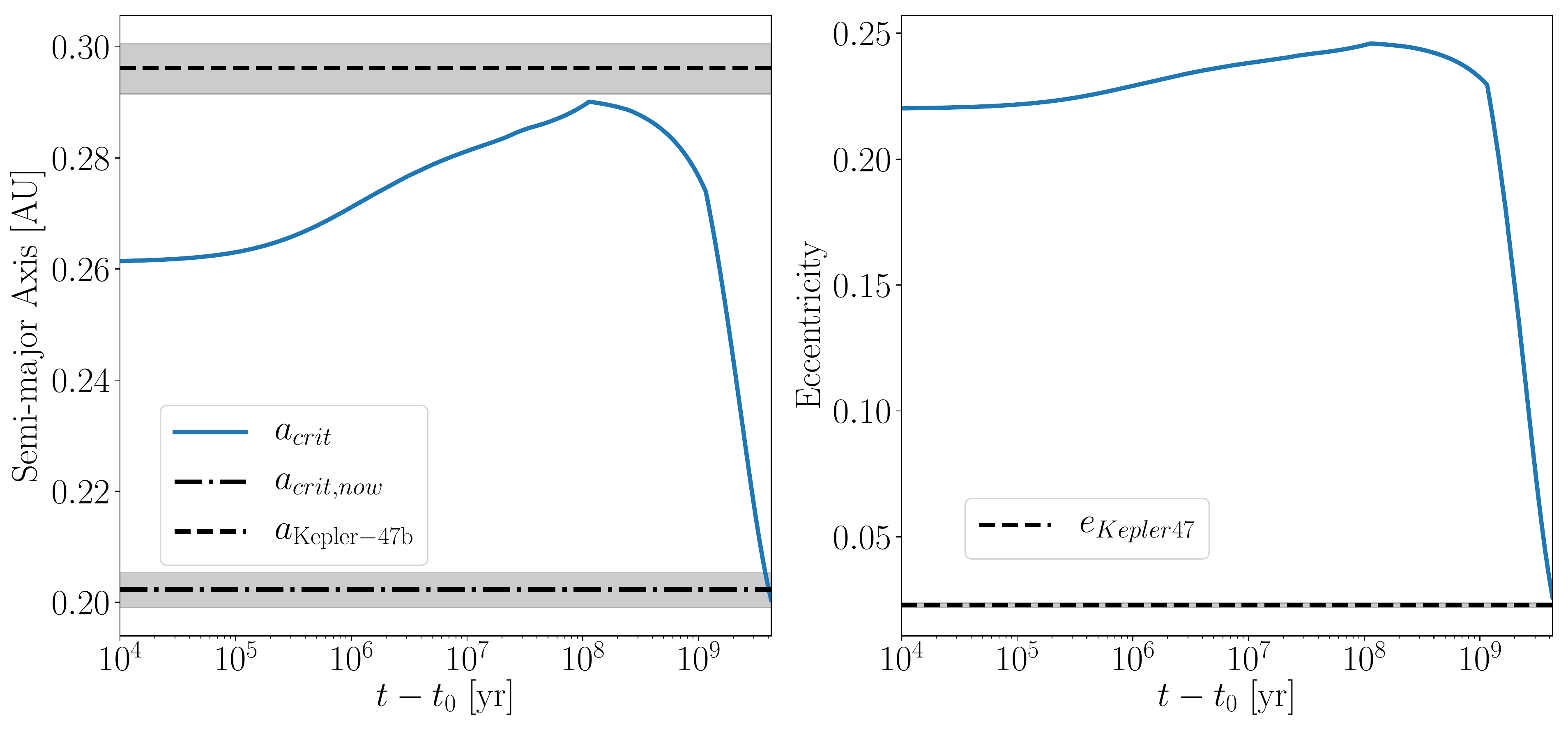}
    \caption{One potential coupled stellar-tidal evolutionary history of Kepler-47. {\it Left:}  $a_{crit}$ evolution.  The blue curve shows a potential past evolution of Kepler-47's $a_{crit}$.  The dashed line represents Kepler-47b's current semi-major axis while the dot-dashed line indicates Kepler-47's observed $a_{crit}$.  The grey regions display the uncertainties on Kepler-47b's semi-major axis and on Kepler-47's $a_{crit}$ from \citet{Orosz2012}. {\it Right:} Eccentricity evolution.  The blue curve displays a potential past evolution of Kepler-47's $e$.  The dashed line and grey regions denote Kepler-47's observed $e$ and the associated uncertainties, respectively.  This potential past evolution of the Kepler-47 achieves an $a_{crit}$ nearly large enough to potentially eject Kepler-47b while also reaching an $a_{crit}$ and $e$ consistent with the observed values \citep{Orosz2012}.  A more rigorous statistical examination of this system is required to constrain its actual past coupled stellar-tidal evolution.}
    \label{fig:kepler47}
\end{figure}

If Kepler-47 did not previously have an additional close-in planet, we examine Kepler-47b, the innermost planet in the Kepler-47 system, which resides at a semi-major axis that is about $1.46 a_{crit}$ for the Kepler-47 binary.  As shown in $\S$~\ref{sec:monte_carlo} and $\S$~\ref{sec:optimistic}, it is plausible for a binary to attain $a_{crit}$ ratios near $1.4$ depending on the initial stellar $P_{rot}$ and stellar tidal Qs.  With Kepler-47's relatively longer orbital period, there was likely an initially larger angular momentum reservoir in the orbit relative to that in the stellar rotations so any transfer would probably not increase the orbital period significantly.  We display such a case of Kepler-47's potential past $a_{crit}$ evolution in Fig.~\ref{fig:kepler47}.  For this simulation, we used the observed stellar masses \citep{Orosz2012}, initial stellar tidal Qs of $5 \times 10^5$, $r_g = 0.45$, an initial $e = 0.22$, and an initial orbital period of 7.5 days.  Note that this case does not represent the precise past evolution of Kepler-47, but instead shows one possible past evolution that is consistent with observations.  In this simulation, given Kepler-47's longer initial orbital period, the binary orbital period and hence $a_{crit}$ grows less via tidal transfer making it less likely to potentially destabilize Kepler-47b.  Under this scenario, Kepler-47b likely formed or migrated to a location far enough away from the central binary to insulate it from the central binary's coupled stellar-tidal evolution, preventing a destabilizing event and making this system compatible with the STEEP process.  Note that accurately modeling Kepler-47's past evolution to assess how the STEEP process could have impacted the circumbinary planetary system requires running a large number of simulations of coupled stellar-tidal evolution and comparing their results with observations.  The results of such a simulation suite could potentially constrain parameters, such as tidal Qs, but that analysis is beyond the scope of this work.

\section{Discussion} \label{sec:discussion}

In this work, we proposed an explanation for the apparent lack of CBPs around isolated binary stars that arises as a natural consequence of short-period binary star evolution.  We showed that binary stars that start with $P_{rot}$ faster than the orbital period transfer angular momentum into the orbit as tides drive the system to a tidally locked state.  The addition of angular momentum to the orbit increases $a$, expanding $a_{crit}$.  Since CBPs tend to preferentially exist near $a_{crit}$, they are then enveloped by the expanding dynamical stability limit and expelled from the system, explaining their observed lack.  We referred to this Stellar Tidal Evolution Ejection of Planets as the STEEP process.  Binary star systems are a product of complex coupled stellar-tidal evolution and their current observed state likely differs appreciably from its past.  Any CBPs that exist around short-period isolated binaries must orbit at large $a_{CBP}$ relative to the binary's $a$ as the past larger region of dynamical stability precludes stable orbits closer to the binary and since the binary's $a$ decays with time post tidal locking due to magnetic braking.  If any future surveys discover new CBPs around short-period isolated binaries, their current state must be understood in the context of the binary's previous evolution.


We examined the dynamical stability of circumbinary planetary systems in which the inner-most planet falls within the region of dynamical instability by running a series of N-body simulations and found that in most cases, at least one planet is ejected.  We performed mock transit observations of such systems after the dynamical stability integration to examine how a potential planetary ejection can impact the detectability any remaining planets.  Most surviving planets' orbits exterior to the ejected planet did not appreciably change after a planet ejection, allowing them to remain detectable via the transit method, although more massive planets tended to scatter away from transiting configurations.  Given the sizable population of short-period binary systems \citep{Kirk2016} around which there are no discovered CBPs, destabilization and subsequent ejection via the STEEP process could have ejected many CBPs contributing to the population of free-floating planets \citep{Veras2012}.  This population could be examined by microlensing surveys \citep[e.g.][]{Sumi2011} and compared with expected free-floating planet population produced by other mechanisms such as planet-planet scattering in single-star systems to gauge its significance.

Our treatment of stellar evolution has its limitations as well, even though we used modern stellar evolution models \citep{Baraffe2015} and magnetic braking laws \citep{Reiners2012,Repetto2014}.  Future examinations of coupled stellar-tidal evolution should model realistic evolving stellar radii of gyration, stellar metallicity effects \citep[e.g.][]{Bolmont2017}, differential rotation \citep[e.g.][]{Lanza2016}, and the effect of binarity on stellar-tidal evolution in order to produce more quantitatively accurate models.  Additionally, future studies should consider directly coupling an N-body code with a coupled stellar-tidal evolution model to robustly model the STEEP process, but we note that such simulations would be quite computationally expensive given the ${\sim}$Gyr timescales of coupled stellar-tidal evolution.

Future examinations of the STEEP process should focus on how coupled stellar-tidal evolution proceeds at large $e$.  If a binary star system tidally locks at $e \gsim 0.2$, it can get captured into a pseudo-synchronous rotation state or a higher order spin-orbit resonance.  We examined the case of an eccentric binary tidally locking into a 3:2 spin-orbit resonance in $\S$~\ref{sec:32}.  We found that at larger $e$, the orbit has less angular momentum so, for given initial stellar $P_{rot}$, tidal transfer of angular momentum to the orbit leads to larger increases in $a_{crit}$ than the synchronous rotation case.  Our model tends to break down at $e \gsim 0.3$ as the CPL model is derived to second order in $e$ and does not resolve tidal locking into pseudo-synchronous rotation like the CTL model does, so future work could examine coupled stellar-tidal evolution using the CTL model.  The precise details of tidal evolution at large $e$, however, is speculative and likely poorly constrained by linear equilibrium tidal models like the CPL and CTL models \citep[e.g.][]{FerrazMello2008,Greenberg2009}.

One effect not modeled by the STEEP process is the impact of mean motion resonances (MMRs) between the CBP and inner binary on the stability of CBP orbits.  \citet{Holman1999} found that the inner-most $n:1$ MMR exterior to $a_{crit}$ produced ``islands" of instability such that CBPs orbiting exterior to $a_{crit}$ near the MMR could still go unstable and be ejected from the system, a finding confirmed by the recent study of \citet{Lam2018}.  Similar to how $a_{crit}$ expands as the binary orbital period increases, the location of the inner-most $n:1$ MMR will extend outward, potentially destabilizing CBPs still residing exterior to $a_{crit}$.  This phenomenon can potentially make the STEEP process more effective at ejecting close-in CBPs and should be examined in future studies.


\subsection{Future Prospects}

The prospect for detecting additional CBPs appears bright as future surveys and algorithmic improvements can potentially increase the known population of CBPs.  Newly discovered CBPs will help characterize the true underlying distribution of $a_{CBP}$ relative to $a_{crit}$ and provide a direct observational test of the destabilization of CBPs via coupled stellar-tidal evolution.  Refined detection algorithms could uncover previously undetected CBPs in both {\it Kepler} and {\it K2} observations.  Future {\it TESS} observations are expected to find approximately 1,100 eclipsing binaries \citep{Sullivan2015} that could host additional CBPs, improving population statistics and helping to settle this remaining issue.  Coupled with the previously-discovered {\it Kepler} eclipsing binaries \citep{Kirk2016}, the eclipsing binaries {\it TESS} will discover will provide a rich dataset that could be used to constrain how coupled stellar-tidal evolution proceeds and could potentially allow for constraints on parameters such as stellar tidal Qs.  Launching between 2022-2024, the {\it ESA}-led {\it PLATO} mission will monitor nearly 1,000,000 stars searching for transits with a focus on low-mass terrestrial planets, potentially discovering new CBPs \citep{Rauer2014}.  Additionally, the {\it Gaia} mission can potentially probe the population of gaseous CBPs and the CBP-binary mutual inclination distribution allowing for comparison with the {\it Kepler} circumbinary population \citep{Sahlmann2015}.  

A key component required to understand the observed CBP population is CBP planet formation.  Although numerous studies have examined planet formation in circumbinary systems \citep[e.g.][]{Alexander2012,Paardekooper2012,Meschiari2012a,Meschiari2012b,Pelupessy2013,Bromley2015,Vartanyan2016}, no previous study has examined the impact of coupled stellar-tidal evolution on young binary stars embedded in a circumbinary disk.  Given that complex disk-binary interactions can lead to significant changes in both the orbit of the binary and the structure of the circumbinary disk \citep[e.g.][]{Fleming2017}, coupled stellar-tidal evolution would necessarily play an important role in that feedback especially since appreciable tidal orbital evolution can occur over the ${\sim}1$ Myr disk lifetime \citep{Haisch2001}.  Although likely computationally non-trivial, accounting for coupled stellar-tidal evolution in simulations of binaries embedded in protoplanetary circumbinary disks could yield new insights into how CBPs form and migrate in circumbinary disks.

In recent years, numerous theoretical modeling efforts have sought to characterize how the presence of two stars impacts the potential habitability of CBPs \citep[e.g.][]{Kane2013,Forgan2014,Popp2017}.  As future studies look to characterize and detect potentially habitable CBPs, we suggest that such efforts should focus on longer-period binaries, those with $P_{orb} \gsim 7.5$ days, as in this regime, the STEEP process is less likely to result in the ejection of close-in CBPs.

As future observations discover and characterize new CBPs, probing the true underlying CBP population, the evolution of the binary stars that host these planets must be well known.    Understanding coupled stellar-tidal evolution in young, short-period binary star systems can provide critical insights into how binaries form and host circumbinary planetary systems and hence provide insights into the observed CBP population.  In this work, we outlined a theoretical framework, the STEEP process, for the long-term evolution of short-period binary stars that provides an explanation for the lack of CBPs around such binary systems.  Future detections or non-detections of CBPs around short-period binaries will provide the best indirect observational test of the STEEP process.

\acknowledgments
The authors wish to thank the anonymous referee for their careful reading of the manuscript and their insightful suggestions that improved the quality of the manuscript.  We also would like to thank Russell Deitrick and Jacob Lustig-Yaeger for constructive conversations that improved the quality of this work.  This work was facilitated though the use of advanced computational, storage, and networking infrastructure provided by the Hyak supercomputer system and funded by the STF at the University of Washington. DPF is supported by an NSF IGERT DGE-1258485 fellowship.  This work was supported by NASA Headquarters under the NASA Earth and Space Science Fellowship Program - Grant 80NSSC17K0482.  DPF, RB, RL, and TRQ acknowledge that this work was supported by the NASA Astrobiology Institute's Virtual Planetary Laboratory under Cooperative Agreement number NNA13AA93A. R.L. acknowledges support from NASA grant NNX14AK26G.

\bibliography{bin_tides}

\end{document}